\documentclass{pasj01}
\onecolumn

\newcommand{\msun}{\thinspace M_\odot}

\newcommand{\dfrac}[2]{{\displaystyle \frac{#1}{#2}}  }

\newcommand{\eqref}[1]{(\ref{#1})}

\def\lesssim{\mathrel{\hbox{\rlap{\hbox{\lower4pt\hbox{$\sim$}}}\hbox{$<$}}}}
\def\gtrsim{\mathrel{\hbox{\rlap{\hbox{\lower4pt\hbox{$\sim$}}}\hbox{$>$}}}}

\begin{document}
\SetRunningHead{Author(s) in page-head}{Running Head}

\title{Significant Gas-to-Dust Ratio Asymmetry and Variation 
in the Disk of HD~142527 and the Indication of Gas Depletion}

\author{Takayuki \textsc{Muto},\altaffilmark{1}
       Takashi \textsc{Tsukagoshi},\altaffilmark{2}
       Munetake \textsc{Momose},\altaffilmark{2}
       Tomoyuki \textsc{Hanawa},\altaffilmark{3}
       Hideko \textsc{Nomura},\altaffilmark{4}
       Misato \textsc{Fukagawa},\altaffilmark{5}
       Kazuya \textsc{Saigo},\altaffilmark{6}
       Akimasa \textsc{Kataoka},\altaffilmark{5,7}
       Yoshimi \textsc{Kitamura},\altaffilmark{8}
       Sanemichi Z. \textsc{Takahashi},\altaffilmark{9}
       Shu-ichiro \textsc{Inutsuka},\altaffilmark{10}
       Taku \textsc{Takeuchi},\altaffilmark{4}
       Hiroshi \textsc{Kobayashi}\altaffilmark{10}
       Eiji \textsc{Akiyama},\altaffilmark{5}
       Mitsuhiko \textsc{Honda},\altaffilmark{11}
       Hideaki \textsc{Fujiwara},\altaffilmark{12}
       and
       Hiroshi \textsc{Shibai}\altaffilmark{13}}
\altaffiltext{1}{Division of Liberal Arts, Kogakuin University, 1-24-2
       Nishi-Shinjuku, Shinjuku-ku, Tokyo 163-8677}
\email{muto@cc.kogakuin.ac.jp}
\altaffiltext{2}{College of Science, Ibaraki University, 2-1-1 Bunkyo, Mito, Ibaraki 310-8512}
\altaffiltext{3}{Center for Frontier Science, Chiba University, 1-33
       Yayoi-cho, Inage-ku, Chiba 263-8522}
\altaffiltext{4}{Department of Earth and Planetary Sciences, Tokyo
       Institute of Technology, Meguro-ku, Tokyo 152-8551}
\altaffiltext{5}{National Astronomical Observatory of Japan, 2-21-1 Osawa, Miaka, Tokyo 181-8588}
\altaffiltext{6}{Department of Physical Science, Graduate School of
       Science, Osaka Prefecture University, 1-1 Gakuen-cho, Naka-ku, Sakai, Osaka 599-8531, Japan}
\altaffiltext{7}{Institute for Theoretical Astrophysics, Heidelberg
       University, Albert-Ueberle-Strasse 2, 69120, Heidelberg, Germany}
\altaffiltext{8}{Institute of Space and Astronautical Science, Japan
       Aerospace Exploration Agency, 3-1-1 Yoshinodai, Chuo-ku, Sagamihara, Kanagawa 252-5210}
\altaffiltext{9}{Astronomical Institute, Tohoku University, 6-3,
       Aramaki, Aoba-ku, Sendai, Miyagi 980-8587}
\altaffiltext{10}{Department of Physics, Graduate School of Science,
       Nagoya University, Furo-cho, Chikusa-ku, Nagoya 464-8601}
\altaffiltext{11}{Department of Mathematics and Physics, Kanagawa
       University, 2946 Tsuchiya, Hiratsuka, Kanagawa 259-1293}
\altaffiltext{12}{Subaru Telescope, 650 North Afohoku Place, Hilo, HI 96720, USA}
\altaffiltext{13}{Graduate School of Science, Osaka University, 1-1
       Machikaneyama, Toyonaka, Osaka 560-0043}


%


\maketitle


\begin{abstract}
 We investigate the dust and gas distribution in the disk around
 HD~142527 based on ALMA observations of dust continuum, 
 $^{13}$CO~$J=3-2$, and C$^{18}$O~$J=3-2$ emission.  
 The disk shows strong azimuthal asymmetry in
 the dust continuum emission, while gas emission is more symmetric.  
In this paper, we investigate how gas and dust are distributed in the
 dust-bright northern part of the disk and in the dust-faint southern
 part.  
 We construct two axisymmetric disk models. 
 One reproduces the radial profiles of the  
 continuum and the velocity moments 0 and 1 of 
 CO lines in the north and the other reproduces those in the south.
 We have found that the dust is concentrated in a narrow ring having 
 $\sim 50~\mathrm{AU}$ width (in FWHM; $w_d=30~\mathrm{AU}$ in our
 parameter definition) located at $\sim 170-200~\mathrm{AU}$ from
 the central star.  The dust particles are strongly concentrated in the
 north.
We have found that 
 the dust surface density contrast between the north and south amounts
 to $\sim 70$.
Compared to the dust, the gas
 distribution is more extended in the radial direction.  
 We find 
 that the gas component extends 
 at least from $\sim 100~\mathrm{AU}$ to
 $\sim 250~\mathrm{AU}$ from the central star, and there should also be
 tenuous gas remaining inside and outside of these radii.
 The azimuthal asymmetry of gas distribution is
 much smaller than dust.
 The gas surface density differs only by a factor of $\sim 3-10$ between
 the north and south.
 Hence, gas-to-dust ratio strongly depends on the location of the disk: 
 $\sim 30$ at the location of the peak of dust distribution in the
 south and 
 $\sim 3$ at the location of the peak of dust distribution in the 
 north.
 Despite large uncertainties, 
 the overall gas-to-dust ratio 
 is inferred to be $\sim 10-30$,
 indicating that the gas depletion may have already been under way.
\end{abstract}

\section{Introduction}

Transitional disks are circumstellar disks having an inner hole of dust
emission, and are considered to be in the evolutionally phase from
gas-rich protoplanetary disks to gas-poor debris ones 
\citep{Strom99, Calvet02, Andrews11}.  This class of
disks have attracted much attention as valuable samples to study the disk
evolution and planet formation processes.
Among a number of transitional disk objects, HD~142527 is a subject of
intense study.  It is a Herbig Fe star \citep{Waelkins96} harboring a
disk with a wide dust cavity with the radius of 
$\gtrsim 100~\mathrm{AU}$
\citep{Fukagawa06,Fujiwara06,Verhoeff11,Rameau12,Casassus12}.  
The stellar mass is $\sim 2.2\msun$ and the age is 5~Myr if we adopt 
$d=140~\mathrm{pc}$ considering 
the association to Sco OB2 \citep{Fukagawa06,Verhoeff11,Mendigutia14}.
Recent observations have revealed the possible existence of low-mass 
companion ($\sim 0.1-0.4~\msun$) at $13~\mathrm{AU}$ 
from the central star \citep{Biller12,Rodigas14}.

The disk around HD~142527 shows several interesting features.
The near infrared scattered light image of the disk shows extended
emission out to $\gtrsim 300~\mathrm{AU}$ and large-scale spiral
features are observed
\citep{Fukagawa06,Casassus12,Canovas13,Rodigas14}, 
hinting that some dynamical activity is taken place 
in the disk~\citep{Casassus12}.
ALMA observations of dust continuum emission show 
significant azimuthal asymmetry with a bright horseshoe-like emission in
the northern part of the disk~\citep{Casassus13,Fukagawa13,Perez14}.  
It is indicated that large grains are concentrated in this northern
region~\citep{Casassus15}.
In the vicinity of the central star, a stream-like features 
in HCO$^{+}~J=4-3$ emission \citep{Casassus13} and a point source of
dust emission \citep{Fukagawa13} are observed.  
HCN~$J=4-3$ and CS~$J=7-6$ emissions are also spatially resolved
with ALMA \citep{vanderPlas14}.

\citet{Fukagawa13} presented the results of ALMA band 7 observations 
of dust continuum and gas emission in $^{13}$CO~$J=3-2$ 
and C$^{18}$O~$J=3-2$.   On the basis of the very bright dust continuum  
emission ($\gtrsim 20~\mathrm{K}$) in the northern part, 
they discussed two possibilities for the disk gas distribution. 
One is that the gas-to-dust mass ratio is less than 100 at least 
in the northern part and significant dust concentration occurs. 
The other is that the gas-to-dust mass ratio is standard value of
100 so the disk gas mass can be high enough for the onset of
gravitational instability.  In either case, the
disk is likely to be in the process of ongoing planet formation.

It is important to pin down the amount of gas and dust 
by means of detailed modeling.
In this paper, we derive the gas and dust distribution based on
ALMA Cycle 0 Band 7 observations of HD~142527 using more 
detailed modeling of gas and dust emission than \citet{Fukagawa13}.  
We derive the dust distribution from the continuum emission and the gas
distribution from the CO emission in order to obtain the gas-to-dust
ratio in the northern and the southern part of the disk.

The paper is constructed as follows.  In section \ref{sec:obs}, we
summarize the results of observations.  
In section \ref{sec:method}, we describe the method of modeling.  
In sections \ref{sec:dustresult} 
and \ref{sec:gasresult}, we present
the modeling results of dust and gas distribution, respectively.  
In section \ref{sec:discussion}, we discuss implications of our
best-fit model.  Section \ref{sec:summary} is for summary.

\section{Observation}
\label{sec:obs}

\subsection{Observation and Data Reduction}
\label{sec:obs_red}

ALMA Band 7 Cycle 0 observations of HD~142527 
 (RA=J2000 15h56m41.9, DEC=J2000 -42d19m23.3)
were carried out 
in the extended array configuration with a maximum baseline of 
about 480~m. The observations consisted of six scheduling 
blocks during the period from June to August 2012. 
The correlator was configured to store dual polarizations in 
four separate spectral windows with 469~MHz of bandwidth and 
3840 channels each, and their central frequencies are 330.588, 329.331, 
342.883 and 342.400~GHz, respectively, to target the molecular lines of 
$^{13}$CO $J=3-2$ and C$^{18}$O $J=3-2$. 
The resultant channel spacing for the lines was 122 kHz, 
corresponding to 0.12~km~s$^{-1}$ in velocity at these frequencies, 
but the effective spectral resolution was lower by a factor 
of around 2 ($\sim$0.2 km~s$^{-1}$) because of Hanning smoothing. 
The continuum data from all the spectral windows were 
aggregated and treated
as a single data set of 336~GHz in central frequency and 
1.8~GHz in bandwidth. The on-source integration after 
flagging aberrant data was 3.0 h. 

Calibration and reduction of the data were made with the 
Common Astronomy Software Applications (CASA) version 3.4, 
in almost the same way as that in \citet{Fukagawa13}. 
Self-calibration was performed for the continuum 
to improve the sensitivity and image fidelity, and 
the final gain solution was also applied to 
$^{13}$CO and C$^{18}$O data.
The only difference from \citet{Fukagawa13} was the visibility 
weighting applied in the final step of the imaging; we adopted 
in this study the Briggs weighting with a robust parameter of 0.5 to 
best recover the weak and extended components of the emission. 
The size in FWHM and the position angle for the major axis of each 
synthesized beam were 
$0\farcs47 \times 0\farcs40 = 60~\mathrm{AU}\times 56\mathrm{AU}$ 
($PA=59.9^{\circ}$), 
$0\farcs50 \times 0\farcs42 = 70\mathrm{AU}\times 59\mathrm{AU}$ 
($PA=57.4^{\circ}$) and 
$0\farcs50 \times 0\farcs42 = 70\mathrm{AU}\times 59\mathrm{AU}$ 
($PA=60.6^{\circ}$), 
for the continuum at 336 GHz, $^{13}$CO and C$^{18}$O,
respectively. Further details on bandpass and gain calibrations are 
described in \citet{Fukagawa13}. 
The rms noise is 0.13 mJy beam$^{-1}$ for the continuum whereas 
it is 6.1 and 8.3 mJy beam$^{-1}$ in the 0.12 km~s$^{-1}$ wide 
channels for the line emission of $^{13}$CO and C$^{18}$O, respectively. 

\subsection{Results of Dust Continuum Emission}
\label{results-cont}

\begin{figure*}
 \begin{center}
  \FigureFile(12cm,6cm){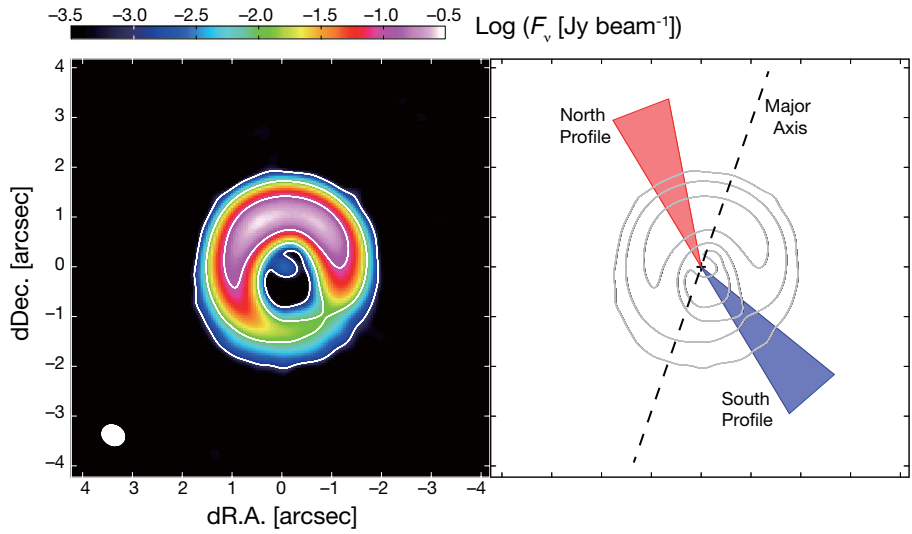} 
 \end{center}
\caption{(Left) 
 Map of the continuum emission at 336 GHz with the Briggs weighting 
with the robust parameter of 0.5.
The synthesized beam, $0\farcs47 \times 0\farcs40$ with 
the major axis $PA=59.9^{\circ}$, is indicated by the white ellipse 
in the bottom left corner.  The contours correspond to 
1, 10 and 100 mJy beam$^{-1}$. The 1$\sigma$ level is 0.13 mJy
 beam$^{-1}$.
(Right)
 The position angle of the major axis is indicated by dashed line.  
 The regions where the azimuthal average is taken
 to obtain the north profiles ($PA=11^{\circ}-31^{\circ}$) and the south
 profiles ($PA=211^{\circ}-231^{\circ}$) are indicated by red and blue
 hatches, respectively.  Contours are the same with the left panel. 
 }
\label{fig:cont}
\end{figure*}

\begin{figure*}
 \begin{center}
  \FigureFile(8cm,6cm){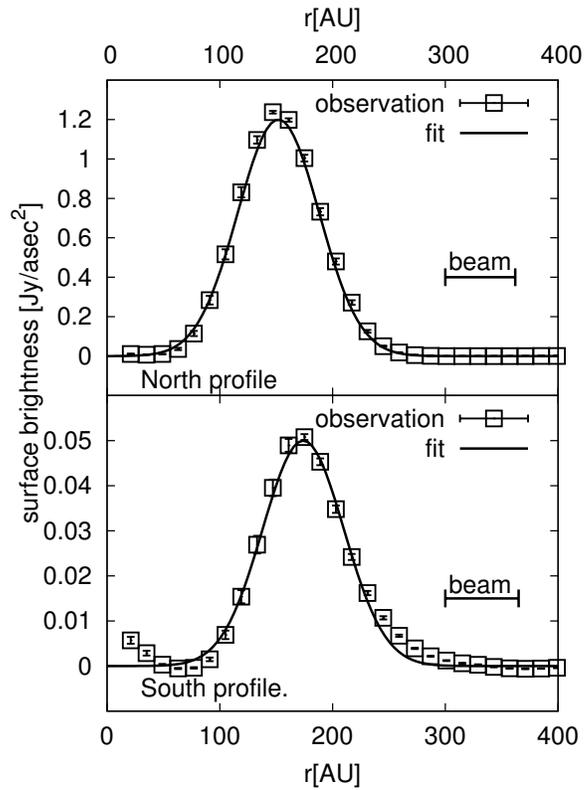} 
 \end{center}
\caption{Radial profiles of surface brightness of the continuum emission 
at 336 GHz in $PA=11-31^{\circ}$ (top panel, squares) and 
$PA=211-231^{\circ}$ (bottom panel, squares). 
Solid lines indicate the best-fit Gaussian function (see Table
 \ref{tab:cont} for parameters).  Error bars indicate 
 the standard deviation after the averaging over 20 degrees in $PA$.}
 \label{fig:radcont}
\end{figure*}

Figure \ref{fig:cont} shows the continuum map, which looks
quite similar to that presented by \citet{Fukagawa13}. 
The outer asymmetric ring as well as an inner 
unresolved component are detected, 
and these are separated by a radial gap. 
The position of the unresolved component coincides with 
the velocity centroid of $^{13}$CO $J=3-2$ (section \ref{results-co}), 
and it is regarded as the stellar position in the following. The radial 
profiles of surface brightness is well described by 
a Gaussian function, and the brightest and faintest of 
their peaks are located at $PA \approx 23^{\circ}$ 
and $PA \approx 223^{\circ}$, respectively (see also 
figure 4 of \citet{Fukagawa13}).  
The averaged surface brightness profiles $I(r)$ in 
$PA=11^{\circ}-31^{\circ}$ and 
$211^{\circ}-231^{\circ}$ are fitted by 
\begin{equation}
I(r) 
 = I_p ~\exp\left[
	     -\frac{(r-r_{0,\mathrm{obs}})^2}{w_{\mathrm{obs}}^2}
	    \right],
 \label{eq:Gauss_obs}
\end{equation}
where $r$ is the angular distance from the star, 
$r_{0,\mathrm{obs}}$ is the peak position and $w_{\mathrm{obs}}$ is the
width of the Gaussian. 
The best-fit profiles are shown in figure \ref{fig:radcont} and the
best-fit parameters are summarized in table \ref{tab:cont}.
As shown in table \ref{tab:cont}, 
the contrast in $I_p$ is $24$ 
between these two position angles. 
In the following, we refer to the averaged profiles in the section of 
$11^{\circ} < PA < 31^{\circ}$ as ``north profile'' and that in 
$211^{\circ} < PA < 231^{\circ}$ as ``south profile'' (see the right
panel of figure \ref{fig:cont}).


\begin{table*}
  \caption{Gaussian parameters (equation \eqref{eq:Gauss_obs}) 
 that fit the radial profiles of dust continuum emission 
 in the brightest and the faintest
 directions.  The parameters $r_0$ and $w$ are given in the unit of AU 
 assuming that the distance to HD 142527 is 140 pc. }\label{tab:cont}
  \begin{center}
    \begin{tabular}{cccc}
      \hline
      $PA$ & $I_p$ [Jy/asec$^{2}$] & $r_{0,\mathrm{obs}}$ [AU] & 
     $w_{\mathrm{obs}}$ [AU]\\  \hline
      $11-31^{\circ}$ (North profile) & 1.2 & 152 & 51 \\
    $211-231^{\circ}$ (South profile) & 0.050 & 174 & 51 \\
      \hline
    \end{tabular}
  \end{center}
\end{table*}


\subsection{Results of $^{13}$CO and C$^{18}$O $J=3-2$ Emission}
\label{results-co}

\begin{figure*}
 \begin{center}
  \FigureFile(15cm,10cm){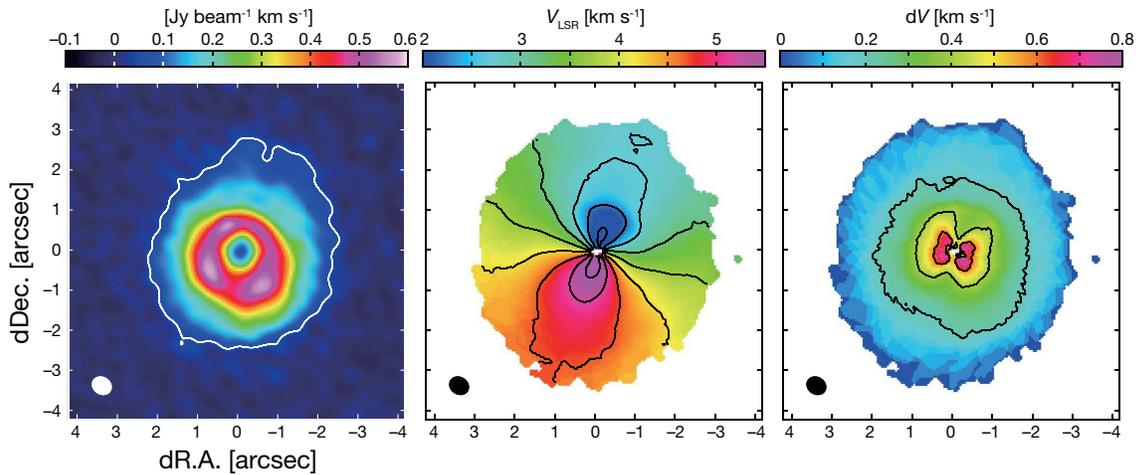} 
 \end{center}
\caption{Moment maps of the $^{13}$CO $J=3-2$ line. 
(left) Moment 0 map, integrated over the velocity 
range of $v_{\mathrm{LSR}} = (0.40-7.00)$ km s$^{-1}$. 
The white contour shows the 5$\sigma$ level (48 mJy beam$^{-1}$ km s$^{-1}$). 
(middle) Moment 1 map, created by the emission above 
the 5$\sigma$ level in channel maps of 0.12 km s$^{-1}$ resolution 
(figure \ref{fig:ch-13co1}-\ref{fig:ch-13co3} in appendix). 
The contours along $PA\approx 71^{\circ}$ are those of the systemic velocity 
($v_{\mathrm{LSR}}$ = 3.7 km~s$^{-1}$), and the contour spacing is 0.5 
km~s$^{-1}$. 
(right) Moment 2 map, created by the emission above 
the 5$\sigma$ level in channel maps of 0.12 km~s$^{-1}$ resolution. 
Contour starts at 0.2 km~s$^{-1}$, and its spacing is 0.2 km~s$^{-1}$.
The synthesized beam, $0\farcs50 \times 0\farcs42$ with 
the major axis $PA=57.4^{\circ}$, is indicated by the ellipse 
in the bottom left corner of each panel.}
\label{fig:13co_mom}
\end{figure*}

\begin{figure*}
 \begin{center}
  \FigureFile(15cm,10cm){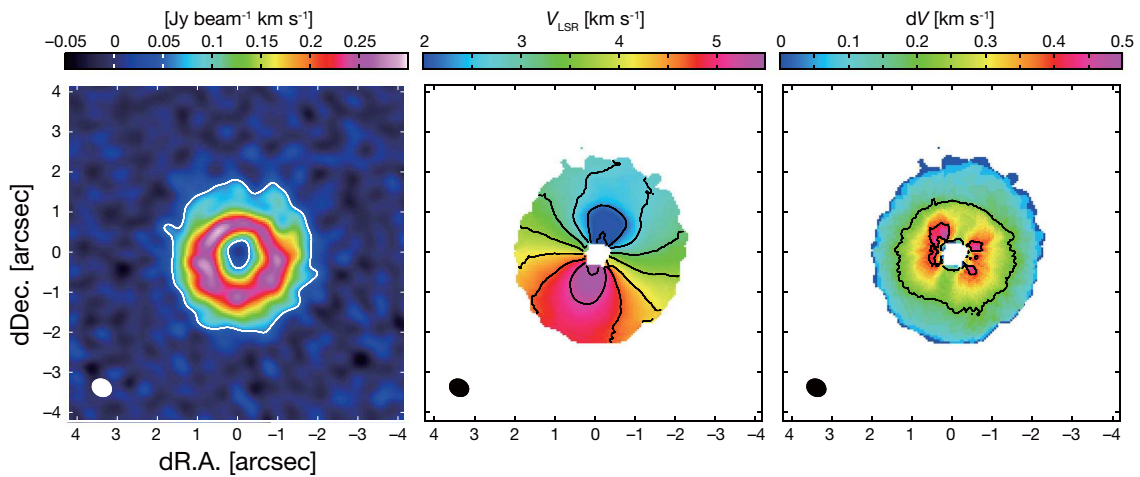} 
 \end{center}
\caption{Moment maps of the C$^{18}$O $J=3-2$ line. 
(left) Moment 0 map, integrated over the velocity 
range of $v_{\mathrm{LSR}} = (1.0-6.40)$ km s$^{-1}$. 
The white contour shows the 5$\sigma$ level (55 mJy beam$^{-1}$ km s$^{-1}$). 
(middle) Moment 1 map, created by the emission above 
the 5$\sigma$ level in channel maps of 0.12 km~s$^{-1}$ resolution 
(figure \ref{fig:ch-c18o1}-\ref{fig:ch-c18o2} in appendix). 
The contours along $PA\approx 71^{\circ}$ are those of the systemic velocity 
($v_{\mathrm{LSR}}$ = 3.7~km~s$^{-1}$), and the contour spacing 
is 0.5~km~s$^{-1}$. 
(right) Moment 2 map, created by the emission above 
the 5$\sigma$ level in channel maps of 0.12 km~s$^{-1}$ resolution. 
Contours at 0.2 and 0.4~km~s$^{-1}$ are shown. 
The synthesized beam, $0\farcs50 \times 0\farcs42$ with 
the major axis $PA=60.6^{\circ}$, is indicated by the ellipse 
in the bottom left corner of each panel.
}\label{fig:c18o_mom}
\end{figure*}

Figures \ref{fig:13co_mom} and \ref{fig:c18o_mom} show the 
moment maps of the $^{13}$CO $J=3-2$ and C$^{18}$O $J=3-2$ emission,
respectively. 
Moments 0, 1, and 2 correspond to the integrated intensity,
intensity-weighted mean velocity, and velocity dispersion,
respectively.  
As shown in the appendix, the emission above $5\sigma$ level 
is detected in $v_{\mathrm{LSR}} = (0.64-7.00)$ km~s$^{-1}$ in 
$^{13}$CO and $v_{\mathrm{LSR}} = (1.24-6.16)$ km~s$^{-1}$ in 
C$^{18}$O. 
The azimuthal asymmetry is weak in the moment 0 map although the
northern part tends to be slightly weaker, possibly due to higher
continuum levels.
Furthermore, the line emission ($^{13}$CO, in particular) is 
clearly detected in the inner regions down to $r \approx 20$~AU 
($0\farcs15$) as well as the outer regions out to $r \approx 400$~AU 
($2\farcs8$). The velocity distribution revealed in 
moments 1 and 2 is consistent with a disk in Keplerian rotation. 
A constant (systemic) velocity of 3.7~km~s$^{-1}$ is found along 
$PA = 71^\circ \pm 2^\circ$ in the moment 1 maps, and this 
is regarded as the direction of the minor axis of the system. 
The position-velocity (P-V) diagram 
along the major axis ($PA = -19^\circ$) is explained well by 
Keplerian rotation with stellar mass of $2.2M_{\odot}$ and 
the inclination angle of $27^{\circ}$, as described in detail 
in the appendix (see also \cite{Fukagawa13,Perez14}). 
We adopt these values for $PA$ of the major axis 
and inclination of the system throughout this paper. 


\begin{figure*}
 \begin{center}
  \FigureFile(12cm,8cm){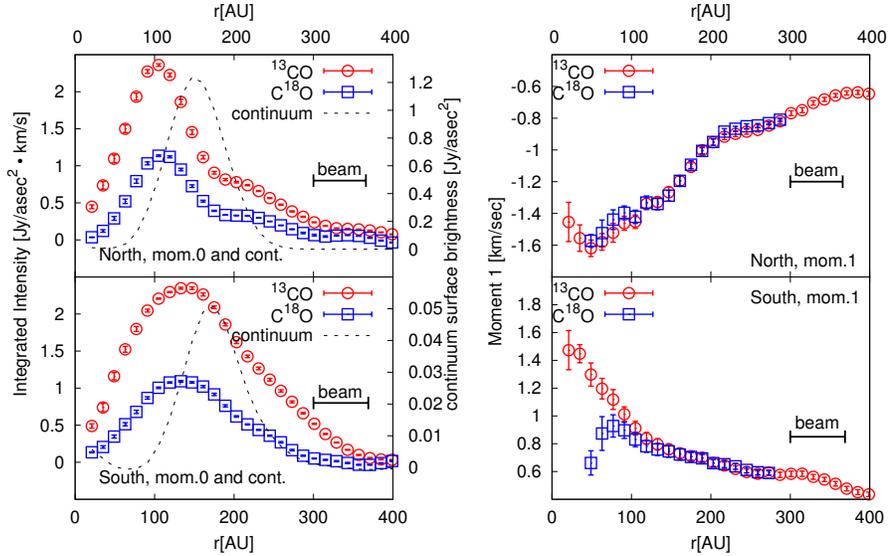} 
 \end{center}
\caption{The north (top row) and the south (bottom row) radial profiles
 of the observed moment 0 (left panels) and moment 1 (right panels) of
 $^{13}$CO (red circles) and C$^{18}$O (blue squares).  
 The systemic velocity of 3.7~km/s is subtracted in calculating the
 moment 1 profiles.
 Error bars
 indicate the standard deviation after the averaging over 20 degrees in
 $PA$.  The radial profiles of the continuum emission is overplotted in
 the moment 0 radial profiles for comparison.}
 \label{fig:gasobs}
\end{figure*}

Figure \ref{fig:gasobs} shows 
the north and south profiles of the moments 0 and 1, which will be 
the main focus of the modeling described in later sections. 
It is clear that the moment 0 profiles of both $^{13}$CO and C$^{18}$O 
are very different from those of dust continuum emission; 
these are more extended than the Gaussian-like dust emission profiles. 
The moment 1 profiles of these two lines 
agree with each other in 
$100\mathrm{AU} \le r \le 280\mathrm{AU}$, indicating 
that both these lines successfully reveal the Keplerian rotation in the 
regions where the emission is detected with a high signal-to-noise
ratio (S/N). 
The moment 1 profile of $^{13}$CO further 
reveals the gas motion down to $r \approx 20$~AU and out to $r \approx 400$~AU. 
As discussed in the subsequent sections, this indicates the
existence of tenuous gas components in these inner and outer regions. 
The $^{13}$CO line has larger moments 0 and 2 in almost all the
positions than the C$^{18}$O.
This is because $^{13}$CO has a larger optical depth at every
velocity channel and hence has a broader line profile 
than C$^{18}$O (see also section \ref{sec:gas_tau})


\section{Method of Modeling}
\label{sec:method}

Our goal is to find density and temperature distribution of
the disk around HD~142527 based on the continuum and CO line emission
profiles described in the previous section.
We construct axisymmetric models 
that can reproduce similar radial brightness profiles in the direction
where dust emission is the brightest (north profile) and the faintest
(south profile) since the model that fully accounts for the azimuthal
asymmetry can be very complex.
This approach is similar to that taken by \citet{Bruderer14} 
in making the model for the disk around Oph IRS 48, which also exhibits
strong azimuthal asymmetry.

\subsection{Dust Distribution Models}
\label{sec:method_dust}

We first derive dust density and temperature distributions 
from the continuum emission  
under the assumption that the disk is in thermal and hydrostatic
equilibrium. 
For simplicity, the gas-to-dust ratio is assumed to be uniform in the
vertical direction and 
the gas temperature is assumed to be the same as dust.  
In other words, 
we have ignored dust sedimentation (e.g., \cite{Dubrulle95})
and temperature difference between gas and dust 
in the low density upper layers 
(e.g., \cite{Kamp04,Nomura05}).
We note that the gas-to-dust ratio can vary in the radial direction, 
which is the main focus of the modeling of 
gas observations 
(see sections \ref{sec:method_gas} and \ref{sec:gasresult}).
The disk is assumed to be heated only by stellar irradiation, 
since viscous heating is less important in the region considered in this
paper.  
The star is assumed to have the effective temperature of 
$T_{\rm eff} = 6250~\mathrm{K}$
and the radius of $R = 3.8~R_{\odot}$ 
\citep{Verhoeff11}.\footnote{The stellar parameters are 
updated in \citet{Mendigutia14}, but the stellar luminosity is 
within the error for the parameters described here.}
We solved the zeroth and first order moment equations 
of the radiative transfer (M1 method, see \cite{Kanno13}). 
We use 226 colors in the wavelength range of 
$0.1~\mu\mathrm{m} \le \lambda \le 3.16~\mathrm{mm}$,  
resulting in the spectral resolution of $\Delta \log \lambda =  0.02$ 
(i.e., $\lambda / \Delta \lambda = 21.7$).   
The computational box covers $30~\mathrm{AU} \le r \le 410~\mathrm{AU}$ 
and $|z| \le 120~\mbox{AU}$ with the spatial resolution of 2 AU in the
cylindrical coordinates. 
The disk is assumed to be symmetric with respect to the midplane.   

\begin{figure*}
 \begin{center}
  \FigureFile(8cm,6cm){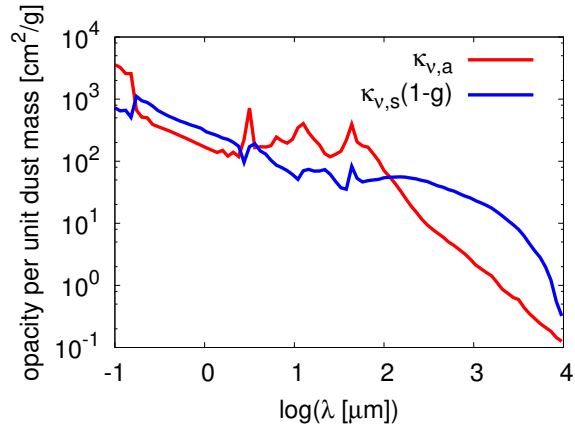} 
 \end{center}
 \caption{The absorption (red) and the effective scattering 
 (blue; see text for definition) 
 coefficients for the dust model used in this paper.}
 \label{fig:dustmodel}
\end{figure*}

The dust is assumed to consist of silicate, carbonaceous grains, and
water ice having the mass fractional abundance of 
$\zeta_{\rm sil}=0.0043$, $\zeta_{\rm carbon}=0.0030$, 
and $\zeta_{\rm ice}=0.0094$, respectively, which are consistent with
solar elemental abundance \citep{Anders89}.  The dust particles are
assumed to have the power-law size
distribution of $\propto a^{-3.5}$ with the maximum size of
$a_{\max}=1~\mathrm{mm}$ \citep{Nomura05}.
Figure \ref{fig:dustmodel} shows the absorption ($\kappa_{\nu,a}$)
and the effective scattering coefficients in
units of cm$^2$ per unit gram of dust.  Here, we define the effective
scattering coefficient as $\kappa_{\nu,s}(1-g)$, where $\kappa_{\nu,s}$
is the scattering coefficient and $g=\langle\cos\theta\rangle$
is the scattering asymmetry factor.\footnote{We have left 
the term $(1-g)$ for consistency with the M1 method.  See 
\citet{Mihalas84,Gonzalez07} for the appearance of 
$(1-g)$ factor in the M1 method.}  
The model with the maximum dust size of 1~{mm} gives the
large value of opacity at sub-mm range \citep{Aikawa06}.  
Consequently, the dust mass evaluated in this paper should be close to
the lowest.
Our opacity at 870~$\mu$m has values of $\kappa_a = 2.9$~cm$^2$/g with 
$\beta\sim 1.2$ at $0.3~\mathrm{mm} \leq \lambda \leq 1~\mathrm{cm}$ and 
$\kappa_s (1-g)=26$~cm$^2$/g per unit dust mass.  The absorption
coefficient at 870~$\mu$m is 20\% smaller than the value adopted by 
\citet{Beckwith90}, which is $\sim 3.5$~cm$^2$/g per unit dust mass.
We also note that the effective scattering coefficient
is large compared to the absorption one in the sub-mm range 
since the maximum dust size is comparable with the wavelength.

The dust surface density is assumed to have the form
\begin{equation}
 \Sigma_d (r) 
  = \Sigma_{d,0} \exp \left[ - \left( \frac{r - r_d}{w_d} \right)
			    ^2 \right]
  \label{eq:dust_model}
\end{equation}
since the radial profile of the continuum emission 
is well approximated by a Gaussian function.
For an assumed set of parameters $(\Sigma_{d,0},r_d,w_d)$, 
we obtain the spatial distribution of dust density $\rho_d(r,z)$,   
temperature $T(r,z)$, and the radiation energy
density $J_{\nu}(r,z)$ for the 226 colors at each grid cell.  

The expected observed surface brightness profiles of dust continuum
emission is obtained by ray-tracing.  
The surface brightness $I_{\nu}$ at frequency $\nu$ is calculated by
solving 
\begin{equation}
 \dfrac{dI_\nu}{ds} = - \rho_d \chi_{\nu} 
  \left[ I_{\nu} - S_{\nu} \right],
\end{equation}
where $s$ is the coordinate along the line of sight, 
$\chi_{\nu}$ is the extinction by absorption and (effective) scattering, 
$\chi_{\nu} = \kappa_{\nu,a}+(1-g)\kappa_{\nu,s}$. 
The source function, $S_{\nu}$, is given by
\begin{equation}
 S_{\nu} = (1-\omega_{\nu}) B_{\nu}(T) + \omega_{\nu} J_{\nu},
  \label{eq:dustsource}
\end{equation}
where 
$\omega_{\nu} = (1-g)\kappa_{\nu,s}/\chi_{\nu}$ 
is the (effective) albedo\footnote{With the dust model of 
$a_{\max}=1~\mathrm{mm}$, the value of 
$\tilde{\omega}_{\nu} = \kappa_s/(\kappa_a+\kappa_s)$ 
and $\omega_{\nu}$ given in the main text 
differs only by $\sim 5\%$ at the wavelengths of interest.}
and $B_{\nu}(T)$ is the Planck function.  
The second term in equation \eqref{eq:dustsource} represents 
the scattered light, 
and the scattering is assumed to be isotropic for simplicity.
The optical depth $\tau_{\nu}$ along the line of sight is given by
\begin{equation}
 \dfrac{d\tau_{\nu}}{ds} = - \chi_{\nu} \rho_d.
\end{equation} 
We compute the model image by using 
$(N_R,N_{\Phi})=(128,128)$ rays covering 
$35~\mathrm{AU}<R<400~\mathrm{AU}$ and $0<\Phi<2\pi$ 
region, where $(R,\Phi)$ are the polar coordinates on the 
sky-plane with the central star at the origin.  We use the inclination
angle of 27$^\circ$, as described in section \ref{sec:obs}

For comparison with observations,    
we convolve the model images with the Gaussian function of the same 
beam size and orientation with the observations. 
We extract the radial surface brightness profiles from the convolved
image and compare them with observations.
We iterate this procedure until 
the given set of parameters $(\Sigma_{d,0},r_d,w_d)$ 
reproduces the observed profiles 
shown in figure \ref{fig:radcont} reasonably well.

\subsection{Gas Distribution Models}
\label{sec:method_gas}

We then derive the gas distribution that accounts for
both moments 0 (integrated intensity) and 1 (intensity-weighted mean velocity) 
radial profiles in the north and south directions.
We have chosen these moments of the line emission 
because they are least affected by beam dilution.  
It should also be noted that 
the observed moment 2 profiles contains the uncertainty coming from
the choice of cutoff levels when producing the moment map from the data.

In later sections, we show that uniform gas-to-dust ratio models do
not reproduce the observations.  
Therefore, we assume that the gas density $\rho_g(r,z)$ is given by
\begin{equation}
 \rho_g(r,z) = \xi(r) \rho_d(r,z),
  \label{eq:gd_def}
\end{equation} 
where $\xi(r)$ represents the gas-to-dust ratio at each radius.  
The gas surface density $\Sigma_{g}(r)$ is then given by
\begin{equation}
 \Sigma_g(r) = \xi(r) \Sigma_d(r)
\end{equation}
In later sections, we look for the forms of $\xi(r)$, or, equivalently,
the form of $\Sigma_g(r)$, that best reproduces the observed radial
profiles of moments 0 and 1.

We assume that the gas rotation is Keplerian at the disk midplane,
\begin{equation}
   v_{\rm rot}(r) 
    = 3.13 \left(\dfrac{r}{200~\mathrm{AU}}\right)^{-1/2}
    [\mathrm{km~s^{-1}}], 
\end{equation}
where $3.13$~km~s$^{-1}$ is the Kepler velocity at 200~AU around a
2.2~$\msun$ star.
Rotation velocity can be slightly different from Keplerian due to, for
example, radial pressure gradient force, 
but the difference is at most of the
order of thermal velocity ($\sim 10\%$ of the Kepler rotation velocity), 
which is hardly observed with current velocity resolution.  
Gas temperature $T(r,z)$ is assumed to be the same as the dust
temperature as mentioned in section \ref{sec:method_dust}. 

The expected brightness of line emission is
calculated with ray-tracing methods by solving
\begin{equation}
 \dfrac{dI_{ul}}{ds} = -\chi_{ul} \left( I_{ul} - S_{ul} \right),
\end{equation}
where $I_{ul}$ is the intensity of the line emission from the upper
state $u$ to the lower state $l$.  
The total extinction $\chi_{ul}$ comes from both dust and gas,
\begin{equation}
 \chi_{ul} = \rho_d \chi_{\nu} 
  + \left( n_l B_{lu} - n_u B_{ul} \right) \phi_{ul} 
  \dfrac{h\nu_{ul}}{4\pi},
  \label{eq:chi_ul}
\end{equation}
and the source function $S_{ul}$ is given by spontaneous emission and
dust emission $S_{\nu}$ given in equation \eqref{eq:dustsource},
\begin{equation}
 S_{ul} = \dfrac{1}{\chi_{ul}} 
  n_u A_{ul} \phi_{ul} \dfrac{h\nu_{ul}}{4\pi} 
  + \dfrac{\rho_d \chi_{\nu}}{\chi_{ul}}S_{\nu} . 
\end{equation}
Here, $n_u$ and $n_l$ are level populations
for the upper and lower state, respectively, $A_{ul}$, $B_{ul}$, 
and $B_{lu}$ are Einstein coefficients, and $\phi_{ul}$ is the line
profile function.  
We note that the scattering of line emission by dust particles is
not included in this work.
Local thermal equilibrium (LTE) is assumed to calculate the level
population,  
which is a valid assumption for lower transition lines of CO
in a protoplanetary disk where typical density is much higher than the
critical densities for these lines~\citep{Pavlyuchenkov07}. 
The fractional abundance of $^{13}$CO and C$^{18}$O is assumed to be  
$9 \times 10^{-7}$ and $1.35 \times 10^{-7}$ \citep{Qi11}.
We have assumed that the line width is determined by thermal
broadening. 

We construct the model channel maps from $-4.5$~km~s$^{-1}$ 
to 4.5~km~s$^{-1}$ with
respect to the systemic velocity with 0.06~km~s$^{-1}$ step
and each model channel map is convolved with the Gaussian beam. 
The model radial profiles of the moments 0 and 1 are then extracted in
the north and south directions for comparison with the observations.
The continuum emission is subtracted on the image by
using line-free channels and the velocity channels showing emission only
below the detection limit are excluded when calculating the moment maps
from the model.  
To confirm the validity of continuum subtraction in the image plane, 
we have also made imaging simulations for several cases in a more 
rigorous way in which the visibilities for the baselines sampled in our 
observations are first generated from the disk model 
with the CASA simulator and then continuum subtraction is made in the
$uv$-plane.  
After imaging with the same procedure 
as that described in section \ref{sec:obs_red}, 
we have confirmed that the resultant line profiles agree with 
those obtained with the image-based continuum subtraction within 
a few percent.

\section{Results of Dust Distribution}
\label{sec:dustresult}

\begin{figure*}
 \begin{center}
  \FigureFile(8cm,14cm){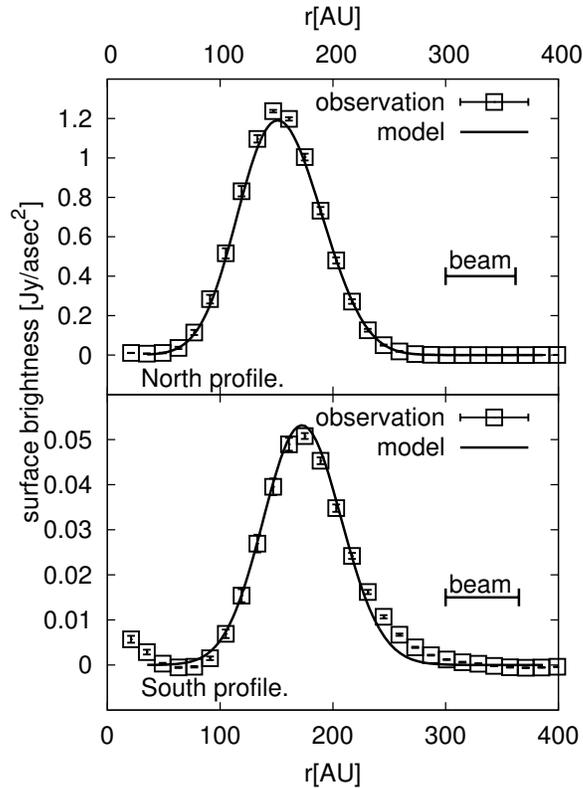}
 \end{center}
\caption{The comparison of continuum emission and the model surface
 brightness for the north (top) and south (bottom) profiles.}\label{fig:cont_model}
\end{figure*}

\begin{figure*}
 \begin{center}
  \FigureFile(15cm,10cm){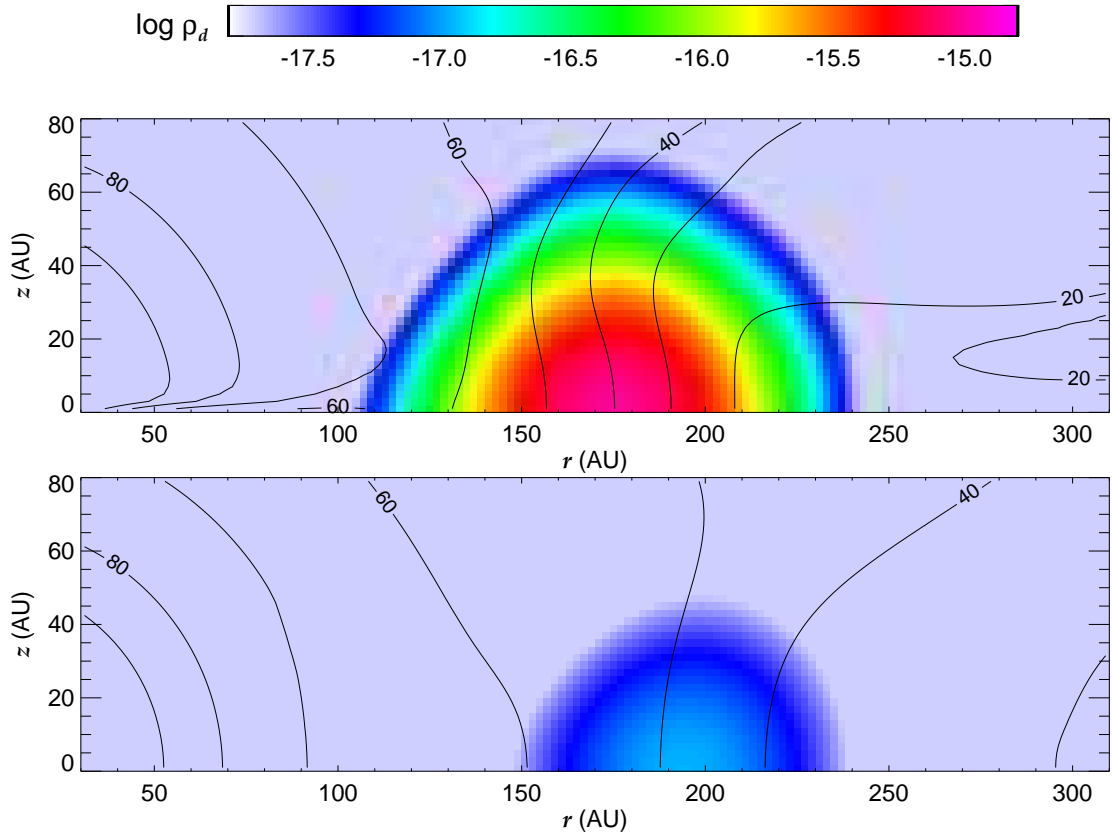} 
 \end{center}
\caption{Dust density and temperature distribution for the best-fit
 models of north (top) and south (bottom) profiles.  Gas
 density is indicated by color in unit of g/cm$^3$ and the temperature
 is shown by contours}\label{fig:rhoT}
\end{figure*}

We have searched for the parameters $\Sigma_{d,0}$, $r_d$, and $w_d$ in
equation \eqref{eq:dust_model} that can reproduce the observed north and
south profiles.  
In the south profiles, the parameter search is straightforward since the
dust emission is optically thin.  In the north, on the other hand, it is
necessary to carefully look at the dependence of surface brightness 
profiles on dust distribution parameters since the dust emission is
optically thick (see section \ref{sec:dust_tau}).  The details of the
parameter search for the north profiles are summarized 
in appendix \ref{sec:app_dustdistribution}.
The best-fit results are summarized in table
\ref{table:dustvalues} 
and figure \ref{fig:cont_model} compares the radial profiles of the dust 
continuum emissions derived from the model and observations.
The best-fit models have the dust density and temperature distributions
shown in figure \ref{fig:rhoT}.
Our derived parameters for
the dust disk (or ring) radius and width is consistent with the results of 
\citet{Verhoeff11}, who show that 
the massive outer disk extends from 130~AU to 200~AU 
from the central star based on their modeling of the SED and
mid-infrared images (and therefore dust emission).

The dust ring emission is only marginally resolved 
in ALMA Band 7 since the full width at half maximum (FWHM) 
of the radial Gaussian function of the model surface density is 
$\sim 2\sqrt{\ln{2}}w_d \sim 50~\mathrm{AU}$ (see table
\ref{table:dustvalues}), which is slightly smaller
than the beam size ($\sim 0\farcs45 \sim 60~\mathrm{AU}$ at 140~pc).  
As a result, the radial width of the observed surface brightness profile
($\sim 2\sqrt{\ln{2}}w_{\rm obs} \sim 85~\mathrm{AU}$ in FWHM; see
table \ref{tab:cont}) is larger than that of the model surface density.
\citet{Perez14} measured the radial width of the dust continuum emission
at 230~GHz to be $0\farcs9$, while at 345~GHz to be $\sim 0\farcs55$.
The difference of the width at different band may be explained by the
effect of 
the convolution by the beam.  The radial width of the continuum emission
is not well resolved.  
The ratio of the radial width of the continuum emission between 230~GHz
and 345~GHz is $\sim 1.6$, which is close to the ratio of the frequency
(and thus the ratio of the beam size) between the two bands.

It is indicated that the amount of dust particles is $\sim 70$ times
more at the peak of the north profiles than at the peak of the south
profile, although the surface brightness contrast of the dust emission 
between the north and the south peak locations is $\sim 24$.  
The difference between the surface density contrast and the surface
brightness contrast is largely due to the fact that the northern part of
the disk is optically thick to dust emission. 
It should also be noted that the
scattered light component in the dust continuum emission is not
negligible in the north profile (see section \ref{sec:optdepth} 
for discussion).

\begin{table*}
\begin{center}
 \begin{tabular}{ccc}
  ~ & North & South \\
  \hline
  \hline
  $\Sigma_{d,0}~[\mathrm{g~cm^{-2}}]$ & 0.6  & $8.45\times 10^{-3}$ \\
  $r_{d}~[\mathrm{AU}]$ & 173 & 196 \\
  $w_{d}~[\mathrm{AU}]$ & 27 & 34  \\
 \end{tabular}
\end{center}
 \caption{Best-fit values for dust distribution.}
 \label{table:dustvalues}
\end{table*}

\section{Results of Gas Distribution}
\label{sec:gasresult}

We now turn our attention to gas distribution.  We first show that 
the models with uniform gas-to-dust ratio 100  
do not reproduce the observed profiles in section \ref{sec:gd100}.  
We then describe in detail how we construct the models for gas
distribution step by step in subsequent subsections.  
Our final results are summarized in section \ref{sec:gassummary}.  

\subsection{Failure of Uniform Gas-to-Dust Ratio Models}
\label{sec:gd100}

\begin{figure*}
 \begin{center}
  \FigureFile(15cm,12cm){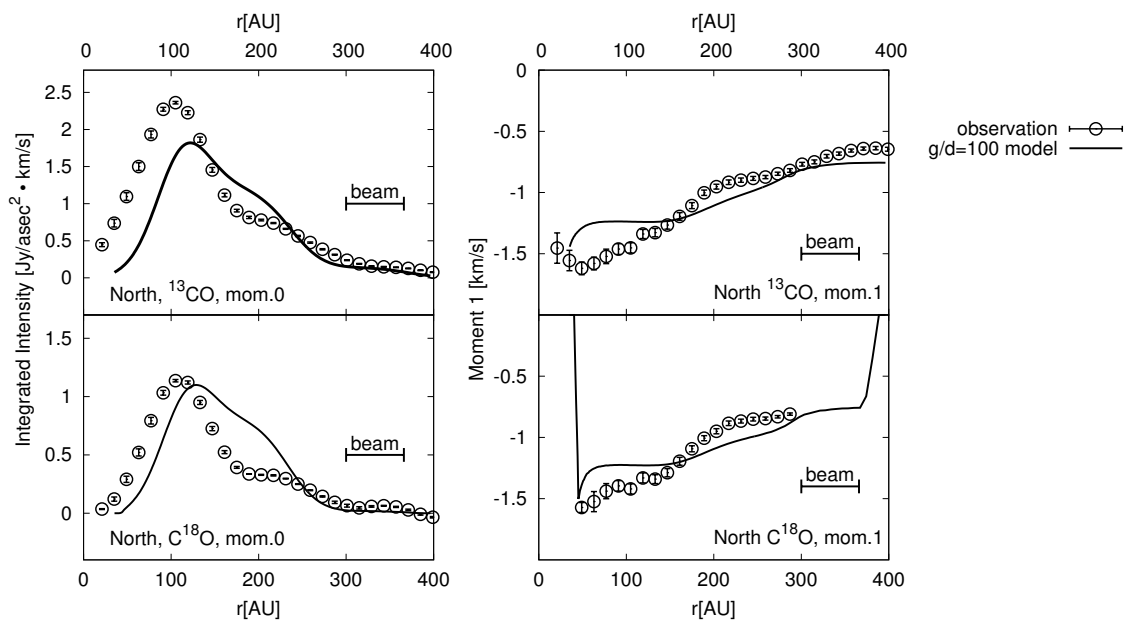} 
 \end{center}
\caption{The radial profiles of moment 0 (left panels) and moment 1
 (right panels) for $^{13}$CO (top row) and C$^{18}$O (bottom row) for
 the north profiles in the case where gas-to-dust ratio is fixed to 100
 everywhere in the disk.}\label{fig:gaussgas_North}
\end{figure*}

\begin{figure*}
 \begin{center}
  \FigureFile(15cm,12cm){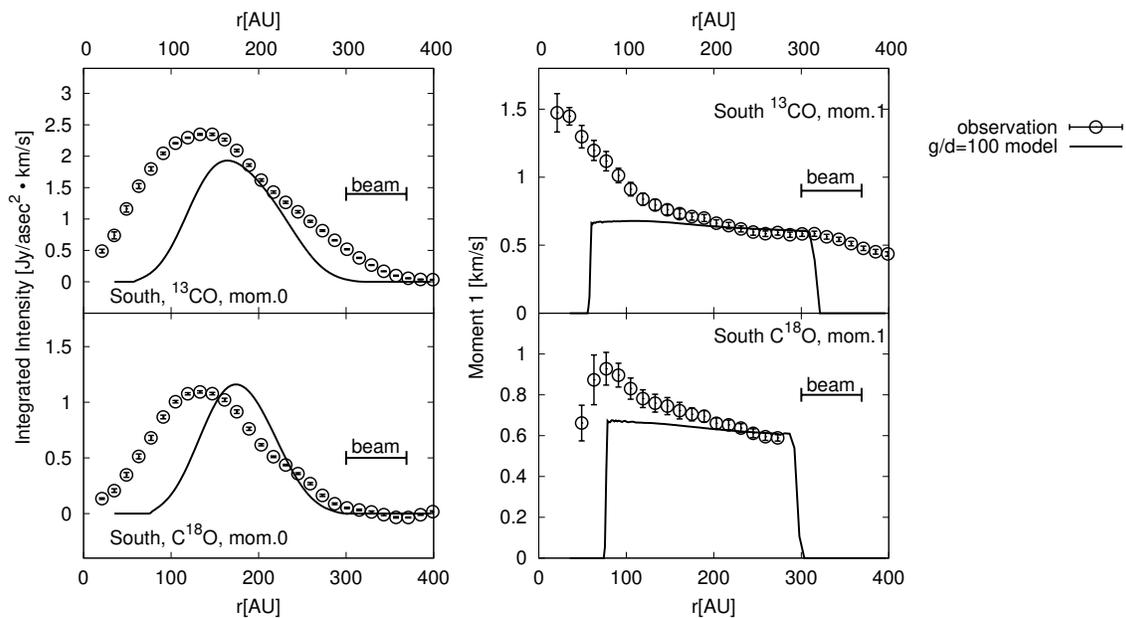} 
 \end{center}
\caption{The same as figure \ref{fig:gaussgas_North} but for the south
 profiles.}\label{fig:gaussgas_South}
\end{figure*}

We first show that the models with uniform gas-to-dust ratio 100,
i.e., $\xi(r)=100$ in equation \eqref{eq:gd_def},
fail to reproduce the observed radial profiles of moments 0 and 1.
In this case, the radial gas surface density profile 
$\Sigma_g(r)=\xi(r)\Sigma_d(r)$ is given by a
Gaussian function as the dust distribution is.  
Figure~\ref{fig:gaussgas_North} shows the radial profiles of 
moments 0 and 1 for the north direction and
figure~\ref{fig:gaussgas_South} shows the same but for the south
direction.  

At inner radii ($r\lesssim 100~\mathrm{AU}$), 
the model profiles of moment 0 show too weak emission compared to
observations and that of moment 1 tends to be 
too slow compared to the observations.
These indicate that there has to be more gas within 100~AU.  
On the other hand, around the peak of the dust
distribution ($r\sim 170~\mathrm{AU}$ for north and 
$r\sim 200~\mathrm{AU}$ for south), 
the model moment 0 is too bright compared to the
observations, especially in the case of C$^{18}$O profiles.  
This suggests that gas-to-dust ratio is smaller than 100 at places 
where dust particles are concentrated.
It is impossible, however, to decrease the amount of gas at 
$\sim 200~\mathrm{AU}$ 
and to increase at inner radii simultaneously under
the assumption of uniform gas-to-dust ratio.  
The dust distribution is already determined 
in section \ref{sec:dustresult}.  
Therefore, we need to consider completely different
radial distribution of the gas from the dust by varying the gas-to-dust
ratio $\xi(r)$ within the disk.

\subsection{Power-Law Gas Profiles}

\begin{table*}
 \begin{center}
  \begin{tabular}{cc}
   \hline
   $\Sigma_0~[\mathrm{g/cm^2}]$ & 0.0845, 0.2325, 0.845, 2.325  \\
   $r_{\rm c}~[\mathrm{AU}]$ & 90, 100, 110, 120  \\
   $r_{\rm out} [\mathrm{AU}]$ & 200, 250, 300, 350  \\
   \hline
  \end{tabular}
 \end{center}
 \caption{Parameters explored for power-law gas distribution given by
 equation \eqref{eq:gasprof_power}.}
 \label{table:params_allcutoff}
\end{table*}

Having found that the uniform gas-to-dust ratio models do not reproduce
the observed profiles of gas emission, 
we now explore models of gas distribution, which is not necessarily
restricted to constant gas-to-dust ratio.  
To acquire consistent results with the dust distribution
calculations described in section \ref{sec:method_dust}, 
we still assume that the gas-to-dust ratio is constant in
the vertical direction {\it at each radius}, 
but it is not constant in the radial direction.
With a trial function of gas surface density $\Sigma_g(r)$, 
it is possible to calculate the gas-to-dust ratio at each radius 
by calculating $\xi(r)=\Sigma_g(r)/\Sigma_d(r)$, 
which is then used to obtain the gas density at each grid cell 
from equation \eqref{eq:gd_def}.  

From the results presented in section \ref{sec:gd100}, 
it is indicated that the gas distribution
is more extended in the radial direction  than dust.  
Therefore, we first try the model
in which the gas surface density profile
$\Sigma_g(r)$ is given by the $r^{-1}$ power-law profile 
at $r_{\rm c}<r<r_{\rm out}$,
\begin{equation}
 \Sigma_g(r) = \Sigma_0 \left( \dfrac{r}{200~\mathrm{AU}} \right)^{-1}
  ~~~~~~~~~~(r_{\rm c}<r<r_{\rm out}),
  \label{eq:gasprof_power}
\end{equation}
and zero otherwise.  Here, $r_{\rm c}$ can be interpreted as the cavity 
radius.
Since constant gas-to-dust ratio models fail to explain the observed
profiles, 
it is necessary to try models with different gas distribution.  The
gas-to-dust ratio is not constant in such cases.
We have first tried several models 
with Gaussian surface density profiles with
different parameters, but we have found that these models fail to
reproduce the observed profiles.  This is primarily due to the fact that
the gas emission extends to large radii, but Gaussian function falls off
too rapidly.  Therefore, we consider power-law distribution of the gas
surface density, 
which is more radially extended than Gaussian profiles.
We choose power-law distribution as a representative model of such
radially extended gas profiles and fix the
power-law exponent to $-1$ in order to reduce the number of free
parameters.  
The choice of $r^{-1}$ power is motivated by the models of 
steady state accretion disk, 
but we do not claim that this is the only possible solution.  
We aim to find one
possible disk model that reasonably reproduces observations 
and discuss its implications for the overall
picture of how the dust and gas are distributed.

We explore the parameter space of 
$(\Sigma_0, r_{\rm c}, r_{\rm out})$ and look for values 
that match the observed moments 0 and 1 profiles 
for the north and south profiles.  
Our strategy is to find first the values of $\Sigma_0$ 
that reasonably matches with the observations at 
$150~\mathrm{AU} \lesssim r \lesssim 200~\mathrm{AU}$,
where dust emission is bright.
Then we look for
the values of $r_{\rm c}$ by investigating the inner radii.
Finally, we search for the values of $r_{\rm out}$ 
by investigating the outer radii.
We show the results one by one in subsequent sections. 
The values of these three parameters explored in our
set of calculations are summarized in
table \ref{table:params_allcutoff}.  

\subsection{The Overall Gas Distribution}
\label{sec:varsig0}

\begin{figure*}
 \begin{center}
  \FigureFile(15cm,12cm){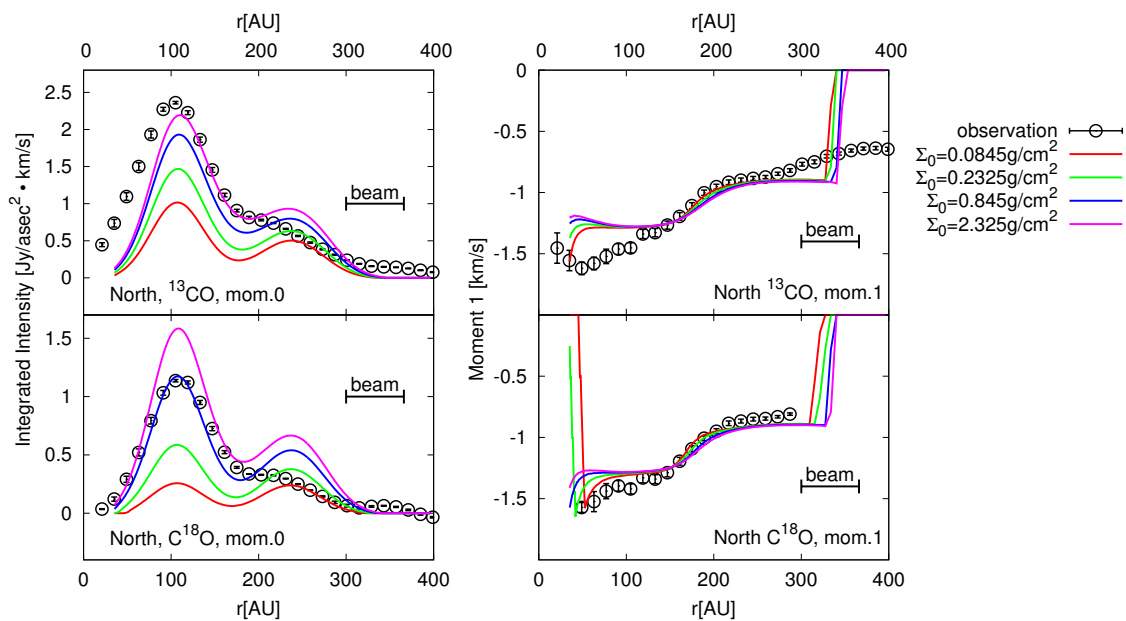} 
 \end{center}
 \caption{The radial profiles of moment 0 (left panels) and moment 1
 (right panels) for $^{13}$CO (top row) and C$^{18}$O (bottom row) 
 for the north direction when the parameter $\Sigma_0$ is varied.}
 \label{fig:varsig0_North}
\end{figure*}

\begin{figure*}
 \begin{center}
  \FigureFile(15cm,12cm){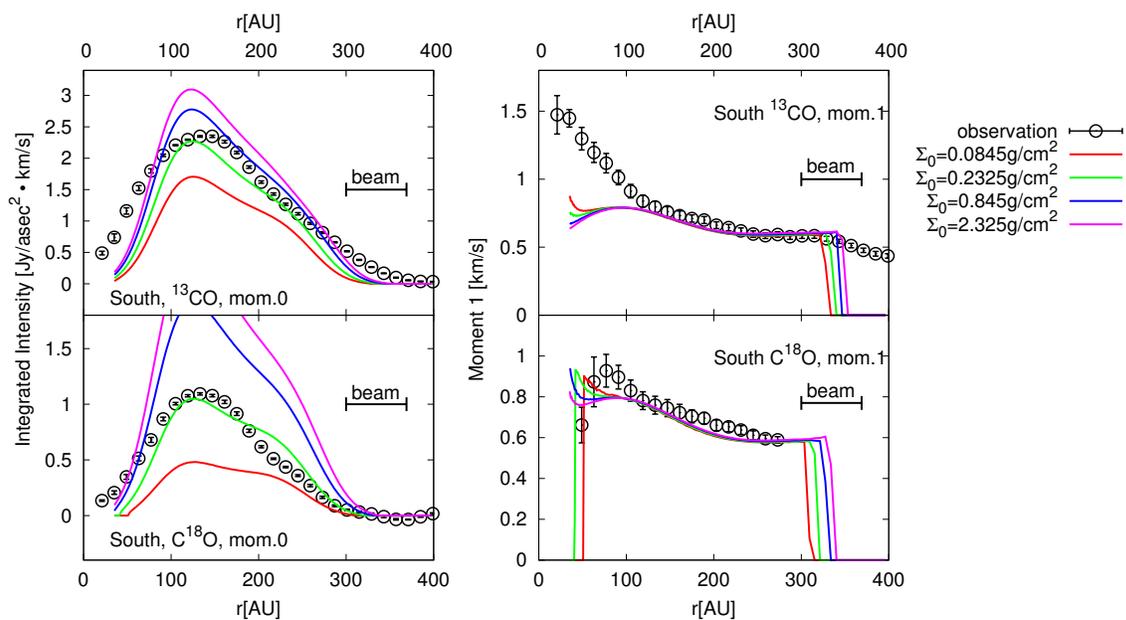} 
 \end{center}
 \caption{The same as figure \ref{fig:varsig0_North} but for the south
 profiles.}
 \label{fig:varsig0_South}
\end{figure*}

We first find the values of $\Sigma_0$ that reasonably
match the observed profiles at 
$150~\mathrm{AU} \lesssim r \lesssim 250\mathrm{AU}$.
For this purpose, we fix 
$(r_{\rm c}, r_{\rm out})=(100~\mathrm{AU}, 280~\mathrm{AU})$ 
for both north and south profiles and we vary $\Sigma_0$ from
$0.0845~\mathrm{g~cm^{-2}}$ 
to $2.325~\mathrm{g~cm^{-2}}$ with a step of a factor of $\sim 3$ (see
table \ref{table:params_allcutoff} for specific parameters).
The lowest value of $\Sigma_0$ 
is chosen in such a way that the gas-to-dust ratio at
the peak of the dust distribution in the south direction is 10.  
Figures \ref{fig:varsig0_North} and \ref{fig:varsig0_South} 
show the results for the north and south profiles, respectively.  

For the north profiles of C$^{18}$O moment 0, the observations fall
between the models with 
$\Sigma_0 = 0.845~\mathrm{g~cm^{-2}}$ 
and $\Sigma_0 = 2.325~\mathrm{g~cm^{-2}}$.
For $^{13}$CO moment 0, the model with 
$\Sigma_0 = 2.325~\mathrm{g~cm^{-2}}$ seems to show better fit, 
but it is
actually difficult to judge which is plausible when the existence of
remaining gas inside the cavity is considered 
(see section \ref{sec:varrin}).  
Note that all the models fail to explain the moment 0 profile 
within the cavity ($r<r_{\rm c}=100~\mathrm{AU}$) 
as the model profiles fall off too rapidly towards inner radii.  
This is because there is no gas emission inside $100~\mathrm{AU}$ in the
model.  
We also point out that the model moment 1 profiles show too slow
velocity at inner radii.  This is because  
moment 1 is calculated to be zero in such gas-free regions 
before we convolve the model with the beam.  The apparent emission
inside the cavity in the model is the result of beam dilution.   
Similar discrepancy between the model and observations can be found at
the outer radii of $r \gtrsim 300~\mathrm{AU}$, while in this case the
beam dilution is not as significant as inner radii.

One prominent feature of the model north profile is the two peaks of the
surface brightness profiles at $r \sim 100~\mathrm{AU}$ and 
$r \sim 250~\mathrm{AU}$, while the gas surface density profile is
smooth.  
These peaks are primarily due to 
the existence of large amount of dust in relatively narrow radial range.
The dust emission is optically thick and is very bright.  The line
emission is partially hidden by the optically thick dust.  
Therefore, the difference in brightness between dust continuum and line
emissions is not very significant.
The line emission at the peak of the dust emission is largely affected
when the continuum emission is subtracted to calculate moment maps.  
The two ``peaks'' of gas emission should be considered as
``one trough'' caused by the subtraction of bright continuum emission.
Although not as prominent as the model profiles, 
it is possible to see this effect in the observed profiles, 
which show a slight dip of the moment 0 at $\sim 170~\mathrm{AU}$ (see
figure \ref{fig:gasobs} and \ref{fig:varsig0_North}). 
We further discuss about these apparent bump and/or trough structures in
section \ref{sec:gasbump}.

For the south profiles, $\Sigma_0=0.2325~\mathrm{g~cm^{-2}}$ models show
reasonable fit to both $^{13}$CO and C$^{18}$O.  However, the
discrepancy between the model moment 1 profiles and the observations at 
inner and outer radii is present, as in the north profile case.
This discrepancy is further studied in sections \ref{sec:varrin} 
and \ref{sec:outerR}.
We note that 
the ``trough'' structure is not very significant in the south profile 
since the continuum emission is weak
and the continuum level is sufficiently low.
Comparing the values of $\Sigma_0$ 
derived for the south profile with that for the north, 
it is indicated that the azimuthal asymmetry of gas distribution
is not as significant as dust distribution.
The azimuthal contrast of gas surface density is only by a
factor of $\sim 3-10$ between the north and south profiles, 
while the peak dust surface density differs by a factor of $\sim 70$.

\subsection{Inner Radius of the Gas and Remaining Gas inside Cavity}
\label{sec:varrin}

We now explore the parameter space for $r_{\rm c}$, the cavity radius of
the gas.  
We fix 
$(\Sigma_0, r_{\rm out}) = (0.845~\mathrm{g~cm^{-2}},280~\mathrm{AU})$ 
and 
$(\Sigma_0, r_{\rm out}) = (0.2325~\mathrm{g~cm^{-2}},280~\mathrm{AU})$  
for the north and south profiles, respectively.
We vary $r_{\rm c}$ from 90~AU
to 120~AU, and see whether we can match the observed profiles within 
$r\lesssim150~\mathrm{AU}$.  Especially, we look for values of 
$r_{\rm c}$ which matches the peak locations of the moment 0 profiles
residing at $r \sim 110~\mathrm{AU}$ in the north profile and 
at $r \sim 140~\mathrm{AU}$ in the south profile.

\begin{figure*}
 \begin{center}
  \FigureFile(15cm,12cm){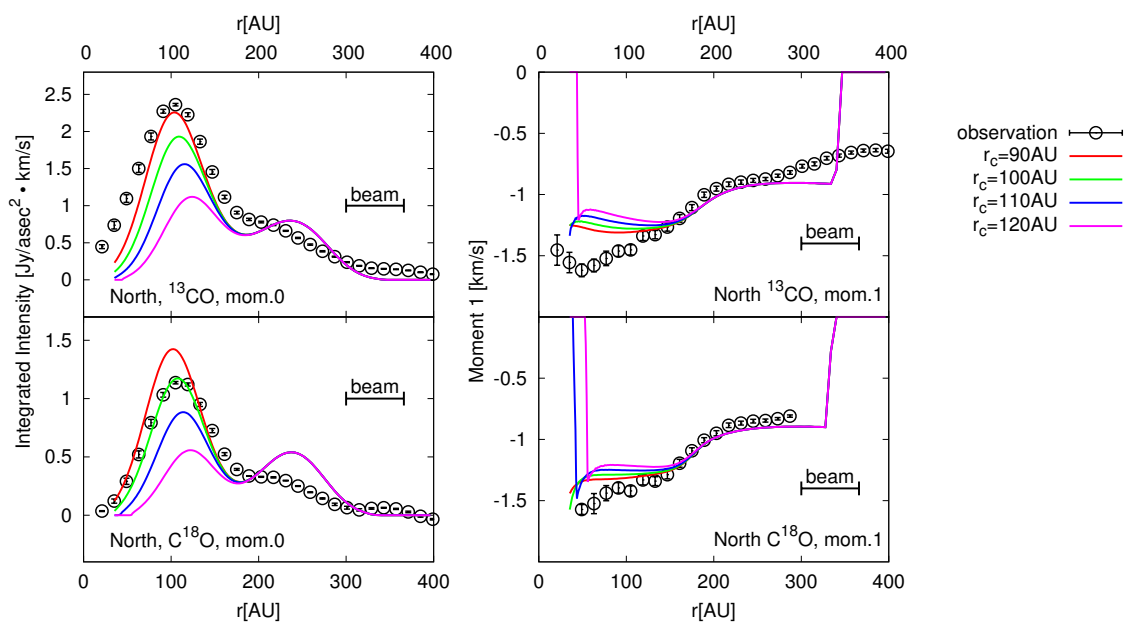} 
 \end{center}
 \caption{The radial profiles of moment 0 (left panels) and moment 1
 (right panels) for $^{13}$CO (top row) and C$^{18}$O (bottom row) 
 for the north direction when the parameter $r_{\rm c}$ is varied.}
 \label{fig:varrin_North}
\end{figure*}

\begin{figure*}
 \begin{center}
  \FigureFile(15cm,12cm){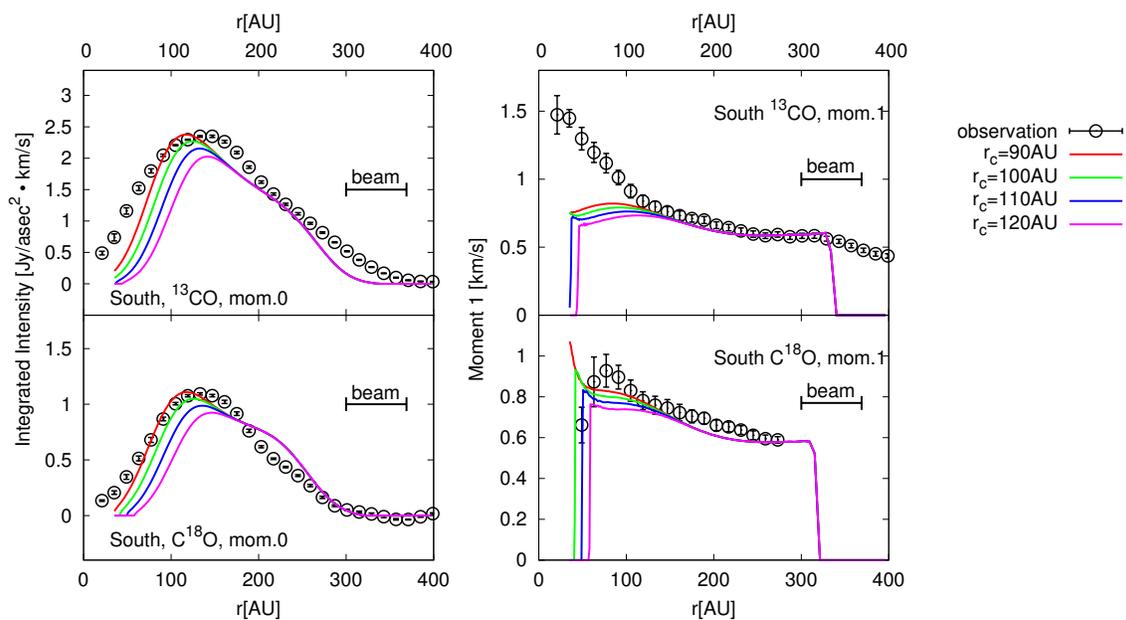} 
 \end{center}
 \caption{The same as figure \ref{fig:varrin_North} but for the south
 profiles.}
 \label{fig:varrin_South}
\end{figure*}

Figures \ref{fig:varrin_North} and \ref{fig:varrin_South} show the
results for the north and south profiles, respectively.  
It is shown that $r_{c}\sim100-110~\mathrm{AU}$ 
can explain the location of the peak in the profiles of moment 0.  
However, all the models give slower velocity in 
moment 1 profile than observations at $r\lesssim 100~\mathrm{AU}$.  
It should also be noted that the model profiles of 
moment 0 fall to zero quickly towards the inner radii 
while the observed profiles,
especially $^{13}$CO south profiles, 
show more gradual decrease.

\begin{figure*}
 \begin{center}
  \FigureFile(15cm,12cm){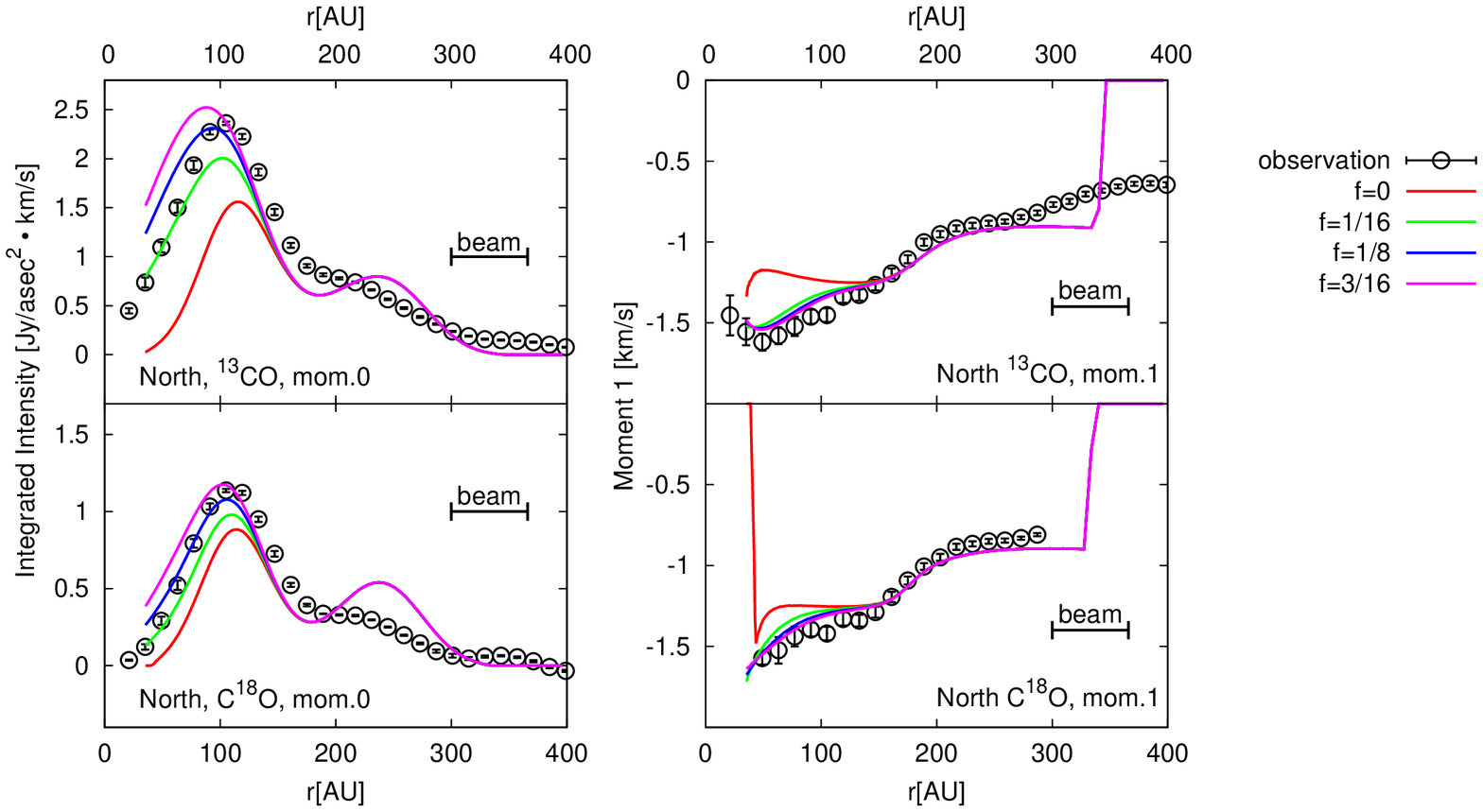} 
 \end{center}
\caption{The radial profiles of moment 0 (left panels) and moment 1
 (right panels) for $^{13}$CO (top row) and C$^{18}$O (bottom row) 
 for the north direction when the parameter $f_{\rm in}$ is varied.}
 \label{fig:varf_North}
\end{figure*}

\begin{figure*}
 \begin{center}
  \FigureFile(15cm,12cm){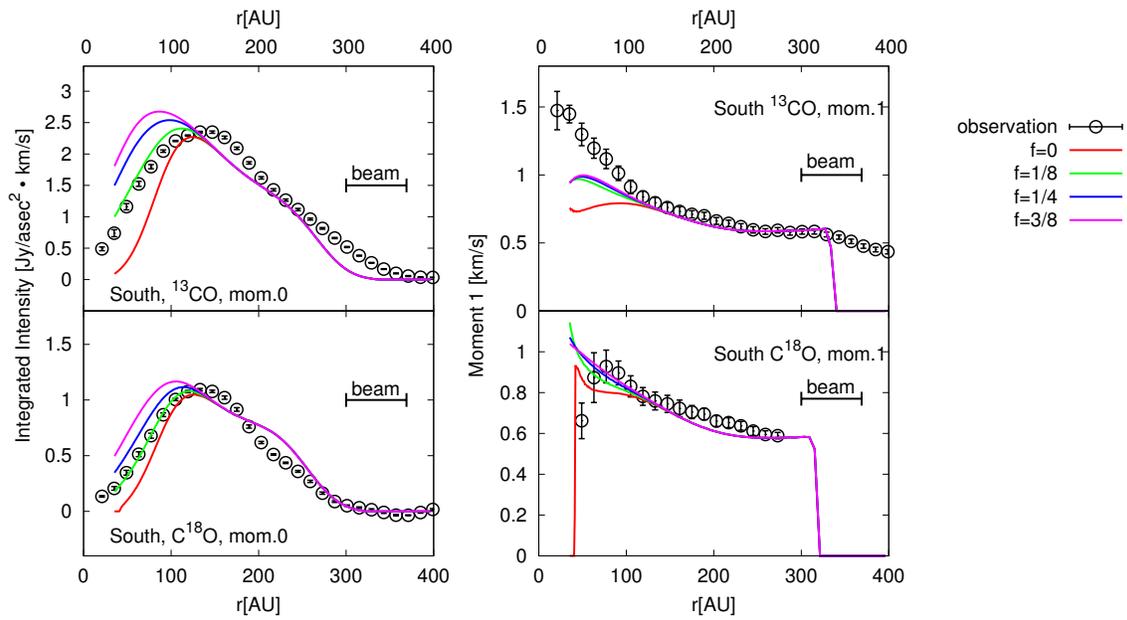} 
 \end{center}
 \caption{The same as figure \ref{fig:varf_North} but for the south
 profiles.}
 \label{fig:varf_South}
\end{figure*}

The discrepancy between the model and observed profiles at inner radii 
can be explained if we assume that 
there is some remaining gas at $r<r_{\rm c}$.  
The existence of gas inside the cavity is also indicated by the
existence of $^{12}$CO emission all the way to the central star
\citep{Casassus13,Rosenfeld14,Perez14}.
To explore the parameters, we assume that
the gas surface density at $r<r_{\rm c}$ is given by
\begin{equation}
 \Sigma_g(r) = f_{\rm in} 
  \Sigma_g(r_c) \left( \dfrac{r}{r_{\rm c}}\right)
  ~~~~~~~~(r<r_{\rm c})
  \label{gassprof_rin}
\end{equation}
while the gas at $r>r_{\rm c}$ is kept the same as equation
\eqref{eq:gasprof_power}.  The parameter $f_{\rm in}$ 
is varied from 0 (complete cavity) to $3/16$ for the north models 
and from 0 to $3/8$ for the south models 
(see table \ref{table:params_cavitygas}). 
Figures \ref{fig:varf_North} and \ref{fig:varf_South} show the results
for the north and south profiles, respectively.  
It is clear that the match between the model and observations 
is better for the models with remaining inner gas.
The values of $f_{\rm in}$ 
that give the best match between the model and 
observations seem to be $\sim 1/8$ 
for the south profile.  For the north model, the best-fit lies somewhere
between $1/16$ and $1/8$.  To keep the parameter search simple, we allow
a factor of $\sim 2$ error here, and use $1/8$ as a representative value.
An important indication from the modeling approach is that similar
gas distribution models (within a factor of $2-3$) can account for the
observations both in the north and south regions of the disk,
which is very different from the case of dust distribution.

We have assumed that the gas surface density increases linearly with
radius within $r<r_{\rm c}$.  However, the functional form of the gas
distribution does not affect the results.  We have checked this by using
a series of models with constant gas surface density at $r<r_{\rm c}$.  
This is partly because the beam size is relatively large 
and the beam dilution effect is significant 
especially when we discuss the gas distribution at inner
radii.  We therefore conclude that it is a robust conclusion that
there should be some remaining gas within the cavity but there is an
uncertainty in the details of how the gas is distributed.
Indeed, \citet{Perez14} have used different functional form for the
models of gas inside the cavity and obtained results consistent with
observations.  The amount of gas within the cavity
in \citet{Perez14}\footnote{The cavity radius is $90$~AU in their model}
is $\sim 1-2 \times 10^{-3} \msun$ while we have the gas mass inside the
cavity being $\sim 5\times10^{-4}\msun$ if we integrate the model with 
$(\Sigma_0, r_{\rm c}, f_{\rm in})=(0.845\mathrm{g/cm^2},110~\mathrm{AU},1/8)$,
which is preferred for the north profile, from $r=0$ to $r=r_c$.  
These results are consistent within an order of magnitude.  
We consider that the difference comes from the fact
that they use lower temperature and smaller line width 
than our model in estimating the gas mass inside the cavity 
(see also section \ref{sec:totalamount}).


\begin{table*}
 \begin{center}
 \begin{tabular}{cc}
  \hline
  $\Sigma_0~[\mathrm{g~cm}$${^2}]$ (north) & 0.845  \\
  $r_{\rm c}~[\mathrm{AU}]$ (north) & 110  \\
  $r_{\rm out} [\mathrm{AU}]$ & 280 \\
  $f_{\rm in}$ (north) & 0, 1/16, 1/8, 3/16 \\
  \hline
 \end{tabular}
 \begin{tabular}{cc}
  \hline
  $\Sigma_0~[\mathrm{g/cm^2}]$ (south) & 0.2325  \\
  $r_{\rm c}~[\mathrm{AU}]$ (south) & 100  \\
  $r_{\rm out} [\mathrm{AU}]$ & 280 \\
  $f_{\rm in}$ (south) & 0, 1/8, 1/4, 3/8 \\
  \hline
 \end{tabular}
 \end{center}
 \caption{Parameters explored for remaining gas inside cavity}
 \label{table:params_cavitygas}
\end{table*}


\subsection{Outer Radius of Dense Gas 
  and Remaining Tenuous Gas at Large Radii}
\label{sec:outerR}

We now turn our attention to the gas distribution at outer radii.  
In the series of models presented in sections \ref{sec:varsig0} and
\ref{sec:varrin}, the outer radius of the gas distribution is fixed to
$r_{\rm out} = 280~\mathrm{AU}$.  As indicated in figures
\ref{fig:varsig0_North}-\ref{fig:varrin_South}, the moment 1 radial
profiles of $^{13}$CO cuts off at $\sim 350~\mathrm{AU}$ while the
observations indicate that there must be some gas at outer radii.

\begin{figure*}
 \begin{center}
  \FigureFile(15cm,12cm){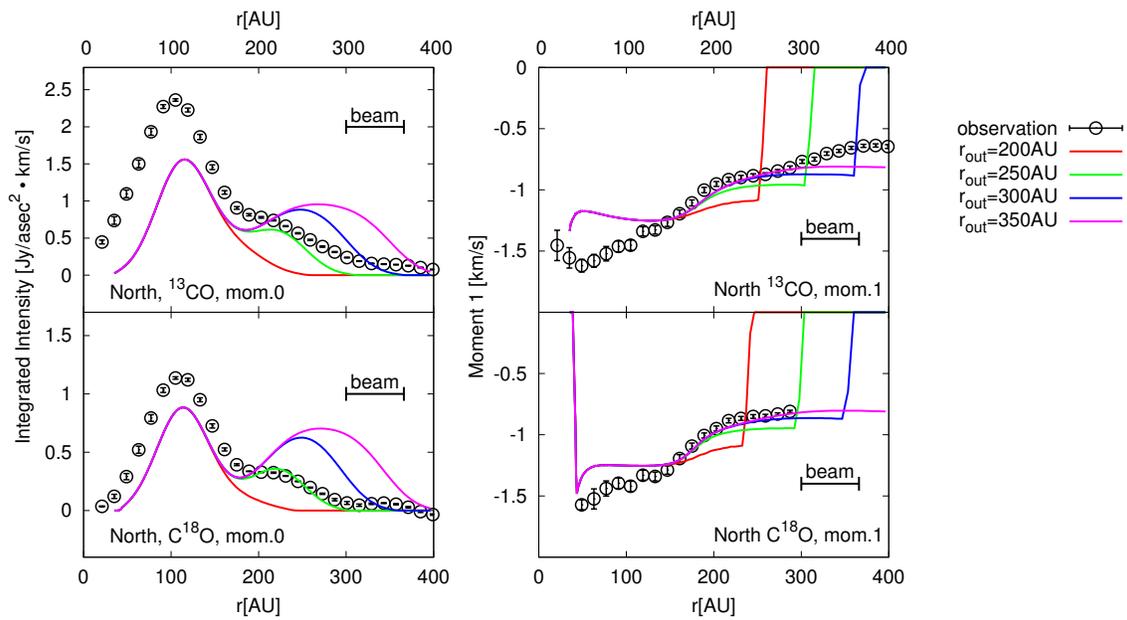} 
 \end{center}
 \caption{The radial profiles of moment 0 (left panels) and moment 1
 (right panels) for $^{13}$CO (top row) and C$^{18}$O (bottom row) 
 for the north direction when the parameter $r_{\rm out}$ is varied.
 For this parameter search, we assume that there is no gas outside
 $r_{\rm out}$.}
 \label{fig:outcut_North}
\end{figure*}

\begin{figure*}
 \begin{center}
  \FigureFile(15cm,12cm){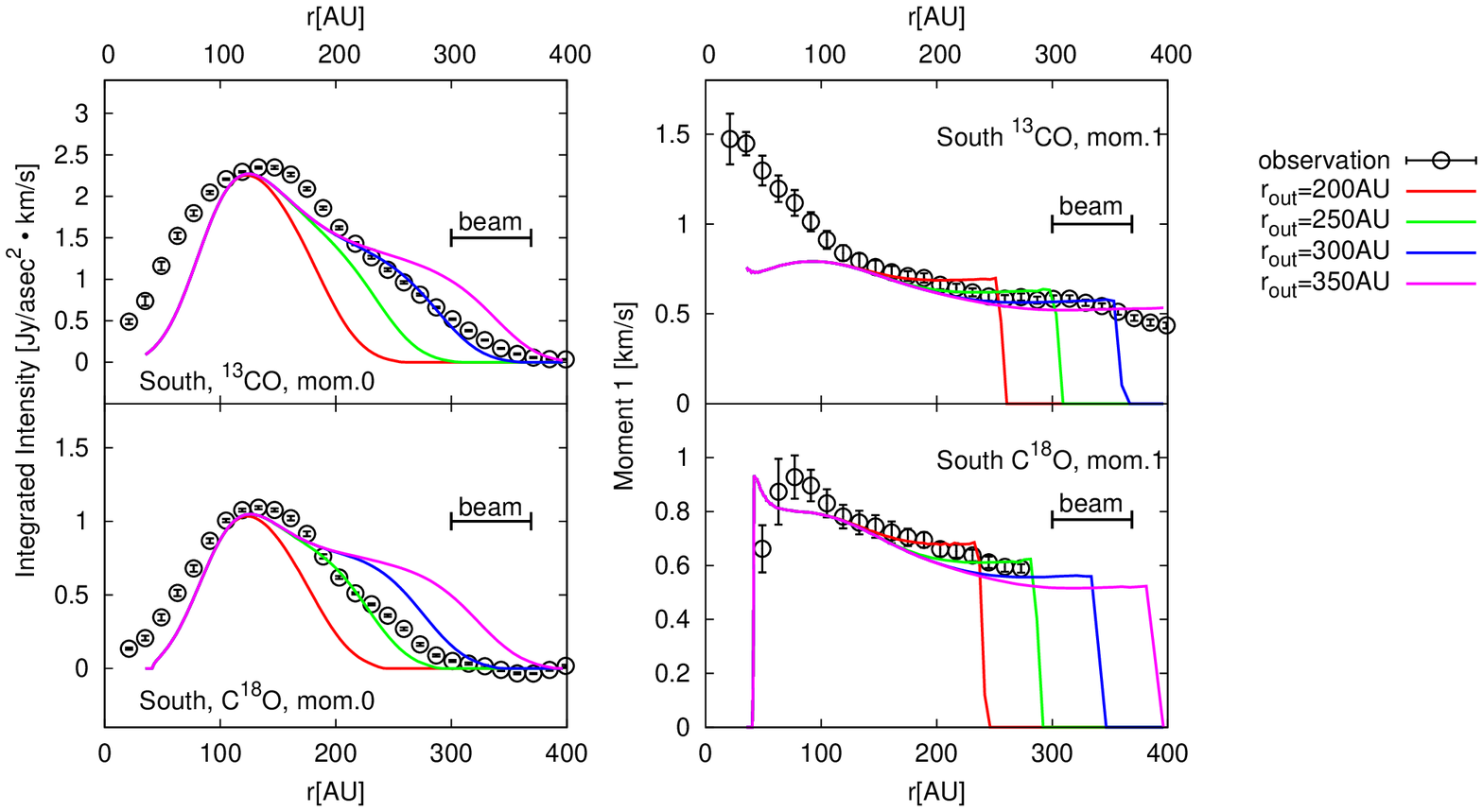} 
 \end{center}
 \caption{The same as figure \ref{fig:outcut_North} but for the south
 profiles.}
 \label{fig:outcut_South}
\end{figure*}

To explore the parameter space, 
we first vary $r_{\rm out}$ while keeping 
$(\Sigma_0,r_{\rm c},f_{\rm in})=(0.845~\mathrm{g/cm^2},110~\mathrm{AU},0)$
and
$(\Sigma_0,r_{\rm c},f_{\rm in})=(0.2325~\mathrm{g/cm^2},100~\mathrm{AU},0)$
for the north and south models, respectively.
We have checked that the details of the choice of 
$r_{\rm c}$ and $f_{\rm in}$ does not affect the brightness profiles at
outer radii. 
Figures \ref{fig:outcut_North} and \ref{fig:outcut_South} show the
results for the north and south profiles, respectively.  
It is shown that the moment 1 profiles 
indeed extend to larger radii as we increase $r_{\rm out}$.  
However, if $r_{\rm out}$ exceeds 250~AU, the models show much larger
values of moment 0 than observed at outer radii.

\begin{figure*}
 \begin{center}
  \FigureFile(15cm,12cm){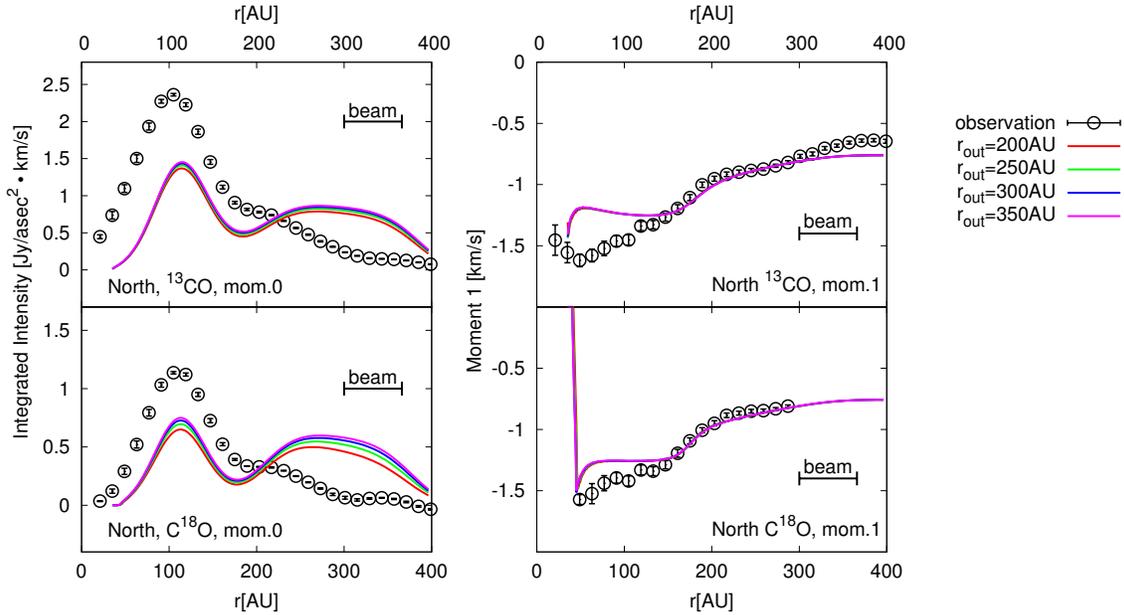} 
 \end{center}
\caption{The radial profiles of moment 0 (left panels) and moment 1
 (right panels) for $^{13}$CO (top row) and C$^{18}$O (bottom row) 
 for the north direction when the parameter $r_{\rm out}$ is varied.
 For this parameter search, we assume that the gas at outer radii
 exhibits exponential cutoff as in equation \eqref{eq:gasprof_exp}.}
 \label{fig:exptail_North} 
\end{figure*}

\begin{figure*}
 \begin{center}
  \FigureFile(15cm,12cm){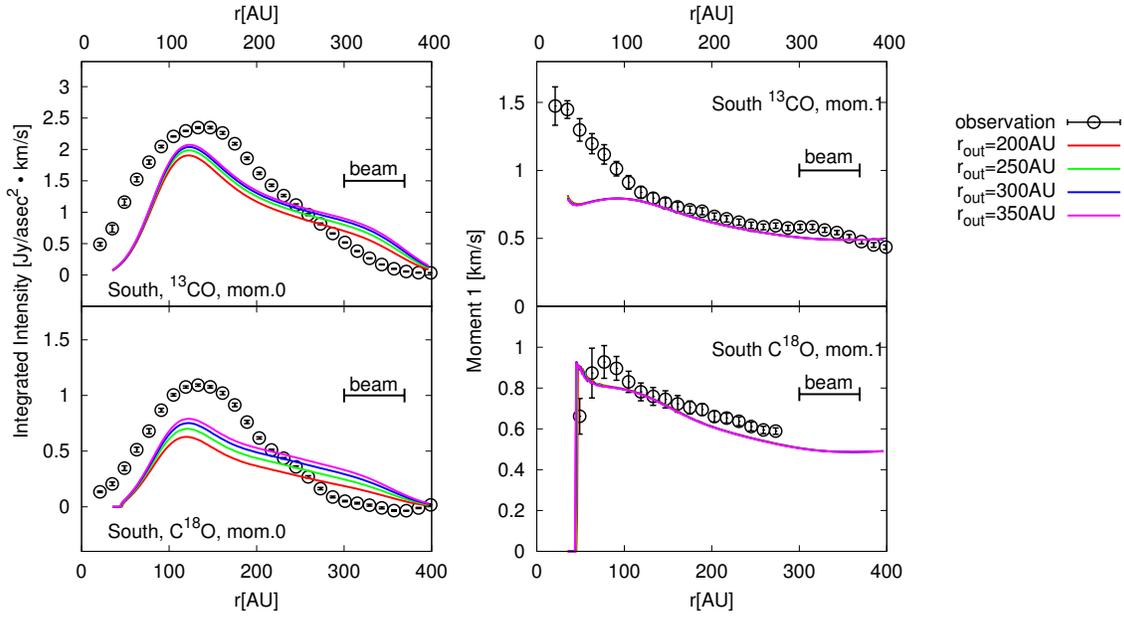} 
 \end{center}
\caption{The same as figure \ref{fig:exptail_North} but for the south
 profiles.}
 \label{fig:exptail_South} 
\end{figure*}

The discrepancy of the moment 0 profiles between the observations and
models indicates that we need to decrease the amount of gas at 
outer radii than currently assumed power-law profiles, especially at
$r>250~\mathrm{AU}$.  At the same time, there should still be some gas 
remaining at outer radii so that moment 1 profiles do not cut off.  
As an alternative series of models, 
we now try commonly used exponential taper model given by
\begin{equation}
  \Sigma_g(r) = \Sigma_0 \left( \dfrac{r}{200~\mathrm{AU}} \right)^{-1} 
   \exp\left( - \dfrac{r}{r_{\rm out}}\right)
  \label{eq:gasprof_exp}
\end{equation}
for $r>r_{\rm c}$.  The results are shown in figures
\ref{fig:exptail_North} and \ref{fig:exptail_South} for the north and
south profiles, respectively.  
Although the model moment 1 profiles show good match 
with observations for both the north and south profiles, 
the models still exhibit too large values for the moment 0
profiles.  This indicates that the gas surface density should decrease
more rapidly than exponential profiles at outer radii. 

To incorporate the rapid decrease of gas surface density at outer radii, 
we finally try a series of model gas surface density profiles 
that cut off at $r_{\rm out}=250~\mathrm{AU}$ 
but there is remaining, tenuous
gas with constant surface density at $r>r_{\rm out}$ all the way to the
outer edge of the computational domain.  
In this framework, the gas surface density 
at $r>r_{\rm out}$ is given by 
\begin{equation}
 \Sigma_{g, \mathrm{out}}(r) 
  = f_{\rm out} \Sigma_g(r_{\rm out})
  ~~~~~(r>r_{\rm out}),
  \label{eq:gasprof_fout}
\end{equation}
where $\Sigma_g(r_{\rm out})$ is the surface density at 
$r=r_{\rm out}$ given by the power-law distribution 
in equation \eqref{eq:gasprof_power} and $f_{\rm out}$ controls the
amount of remaining gas at outer radii.  The parameters explored are
given in table \ref{table:params_tenuousgas}.

\begin{figure*}
 \begin{center}
  \FigureFile(15cm,12cm){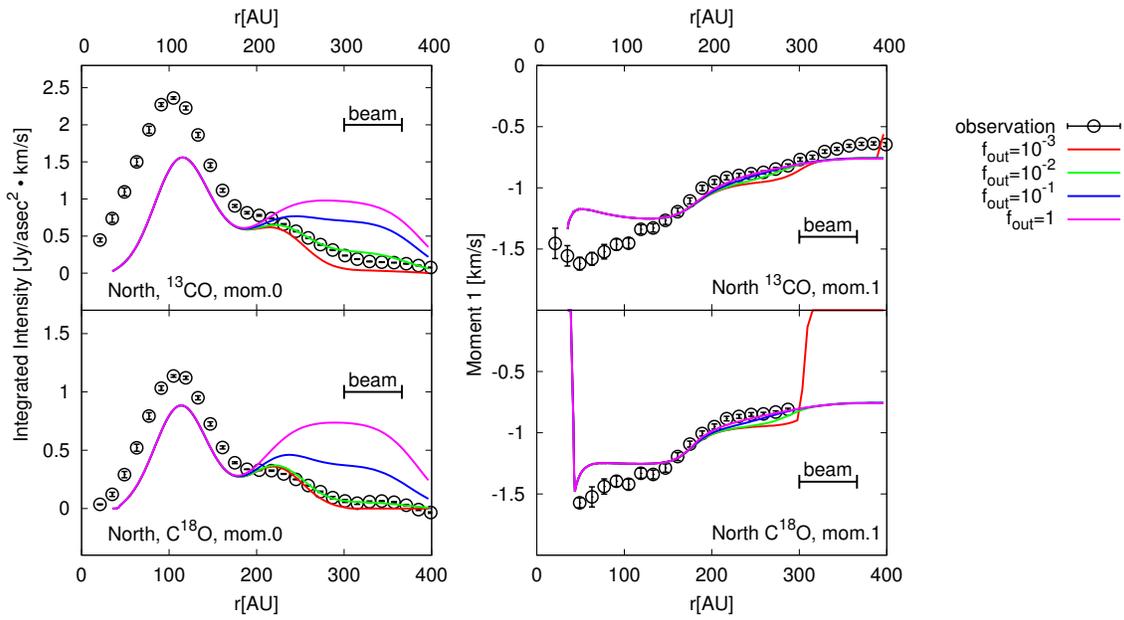} 
 \end{center}
 \caption{The radial profiles of moment 0 (left panels) and moment 1
 (right panels) for $^{13}$CO (top row) and C$^{18}$O (bottom row) 
 for the north direction when the parameter $f_{\rm out}$ is varied.
 For this parameter search, we assume that the gas at outer radii
 exhibits a cutoff at 250~AU, but there is still some remaining gas
 outside the cutoff radius as in equation \eqref{eq:gasprof_fout}.}
 \label{fig:outremain_North}
\end{figure*}

\begin{figure*}
 \begin{center}
  \FigureFile(15cm,12cm){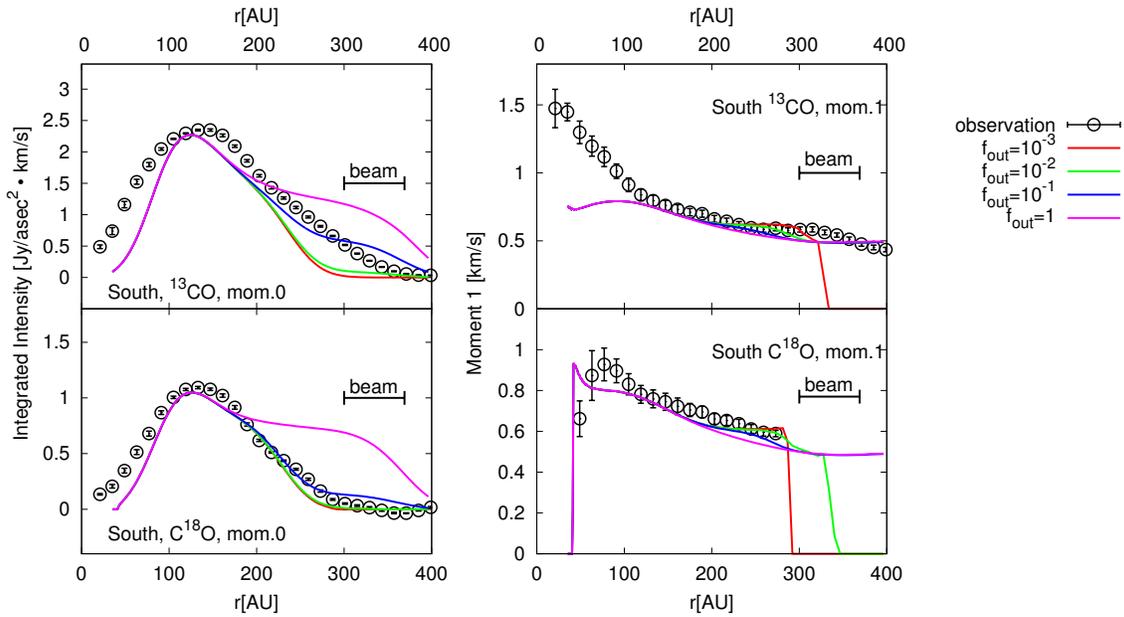} 
 \end{center}
 \caption{The same as figure \ref{fig:outremain_North} but for the south
 profiles.}
 \label{fig:outremain_South}
\end{figure*}

Figures \ref{fig:outremain_North} and \ref{fig:outremain_South} show the
results for the north and south profiles, respectively.
For the north profiles, the model profiles at 
$r\gtrsim 250~\mathrm{AU}$ show reasonable match with observations 
when $f_{\rm out} \sim 10^{-2}$.  For the south profiles, 
$f_{\rm out} = 10^{-1}$ models show agreement 
at $r\lesssim 300~\mathrm{AU}$ while they show brighter emission 
than observations at the outermost radii.  
The moment 0 profiles of the model with $f_{\rm out}=10^{-2}$ 
is fainter than observations all the way in the outer radii.  
Therefore, the actual values of 
$f_{\rm out}$ may vary as a function of radius 
for the south profiles, but the overall value of $f_{\rm out}$ 
may be around $\sim 10^{-1}$ to $\sim 10^{-2}$.
Finally, we briefly note that 
the appearance of the ``trough'' at $r\sim 170~\mathrm{AU}$ 
in the north profile (see discussions in section 
\ref{sec:varsig0}) is different when we use different profiles 
at $r \gtrsim r_{\rm out}$.  
This is because the change in gas surface density at outer radii
(within the parameters explored in this section) mainly affects 
the moment 0 profiles at $r\gtrsim 200~\mathrm{AU}$. 
The strength of the second (apparent) bump at 
$r\sim 250~\mathrm{AU}$ is therefore affected.
 
\begin{table*}
 \begin{center}
 \begin{tabular}{cc}
  \hline
  $\Sigma_0~[\mathrm{g/cm^2}]$ (north) & 0.845  \\
  $r_{\rm c}~[\mathrm{AU}]$ (north) & 110  \\
  $r_{\rm out} [\mathrm{AU}]$ & 250 \\
  $f_{\rm out}$ & $1$, $10^{-1}$, $10^{-2}$, $10^{-3}$ \\
  \hline
 \end{tabular}
 \begin{tabular}{cc}
  \hline
  $\Sigma_0~[\mathrm{g/cm^2}]$ (south) & 0.2325  \\
  $r_{\rm c}~[\mathrm{AU}]$ (south) & 100  \\
  $r_{\rm out} [\mathrm{AU}]$ & 250 \\
  $f_{\rm out}$ & $1$, $10^{-1}$, $10^{-2}$, $10^{-3}$ \\
  \hline
 \end{tabular}
 \end{center}
 \caption{Parameters explored for tenuous gas outside}
 \label{table:params_tenuousgas}
\end{table*}

\subsection{Summary of Gas Distribution}
\label{sec:gassummary}

We have looked for the gas distribution models 
that reproduce the observed moments 0 and 1
profiles of $^{13}$CO and C$^{18}$O.  We have shown that the 
gas-to-dust ratio is not constant in the disk and that the gas
distribution should be more extended in the radial directions 
than the dust distribution.  
We have seen that there should be some remaining gas inside 
$\sim 100~\mathrm{AU}$ and some tenuous gas at 
$r\gtrsim 250~\mathrm{AU}$. 

To be more specific, we have assumed 
that the model surface density of gas is given by
\begin{eqnarray}
 \Sigma_g(r) =
  \left\{
   \begin{array}{cc}
    f_{\rm in} \sigma(r_{\rm c}) \left(\dfrac{r}{r_{\rm c}} \right) &
     (r<r_{\rm c}) \\
     \sigma(r) & (r_{\rm c}<r<r_{\rm out}) \\
     f_{\rm out} \sigma(r_{\rm out}) & (r>r_{\rm out})
   \end{array}
  \right. ,
  \label{eq:finalmodel}
\end{eqnarray}
where $\sigma(r)$ is given by the power-law distribution,
\begin{equation}
 \sigma(r) = \Sigma_0 \left(\dfrac{r}{200~\mathrm{AU}}\right)^{-1}.
\end{equation}
There are five control parameters: 
$\Sigma_0$ that determines the overall amount gas surface density, 
$r_{\rm c}$ that determines the inner cutoff (cavity) radius, 
$f_{\rm in}$ that determines the amount of gas within the cavity,
$r_{\rm out}$ that determines the outer radius of the disk, and
$f_{\rm out}$ that determines the amount of tenuous gas at large radii.
The values of these parameters that give reasonable fit to
observations are given in table~\ref{table:finalvalues}.
Hereafter, we call these best-fit models ``reference models''.  

\begin{figure*}
 \begin{center}
  \FigureFile(15cm,12cm){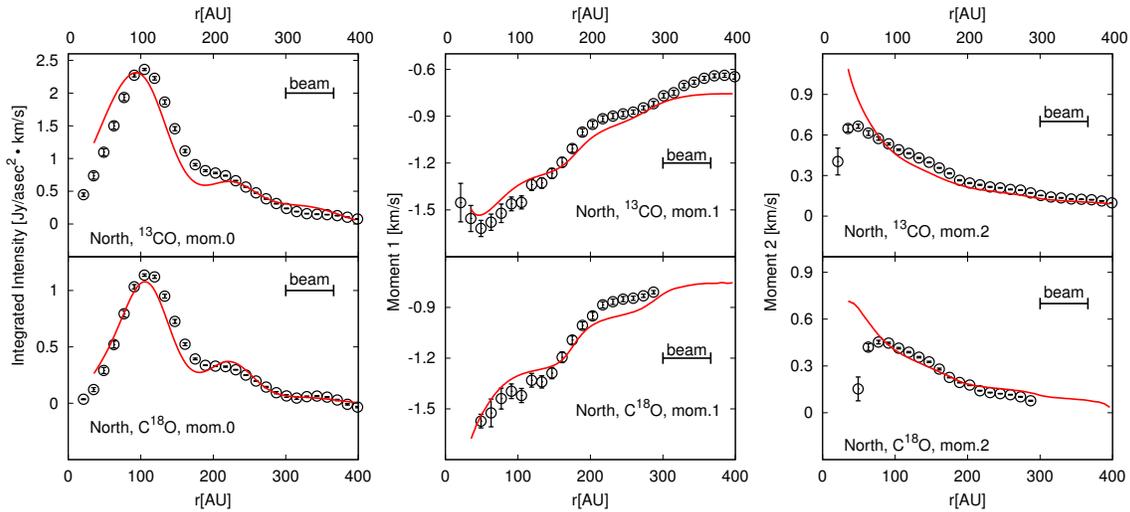} 
 \end{center}
 \caption{North radial profiles of moment 0 (left panels), 
 moment 1 (middle panels), and moment 2 (right panels) of $^{13}$CO (top
 row) and C$^{18}$O (bottom row) for the reference 
 model given by equation \eqref{eq:finalmodel}.   
 The values of the model parameters are given 
 in table \ref{table:finalvalues}.}
 \label{fig:final_North}
\end{figure*}

\begin{figure*}
 \begin{center}
  \FigureFile(15cm,12cm){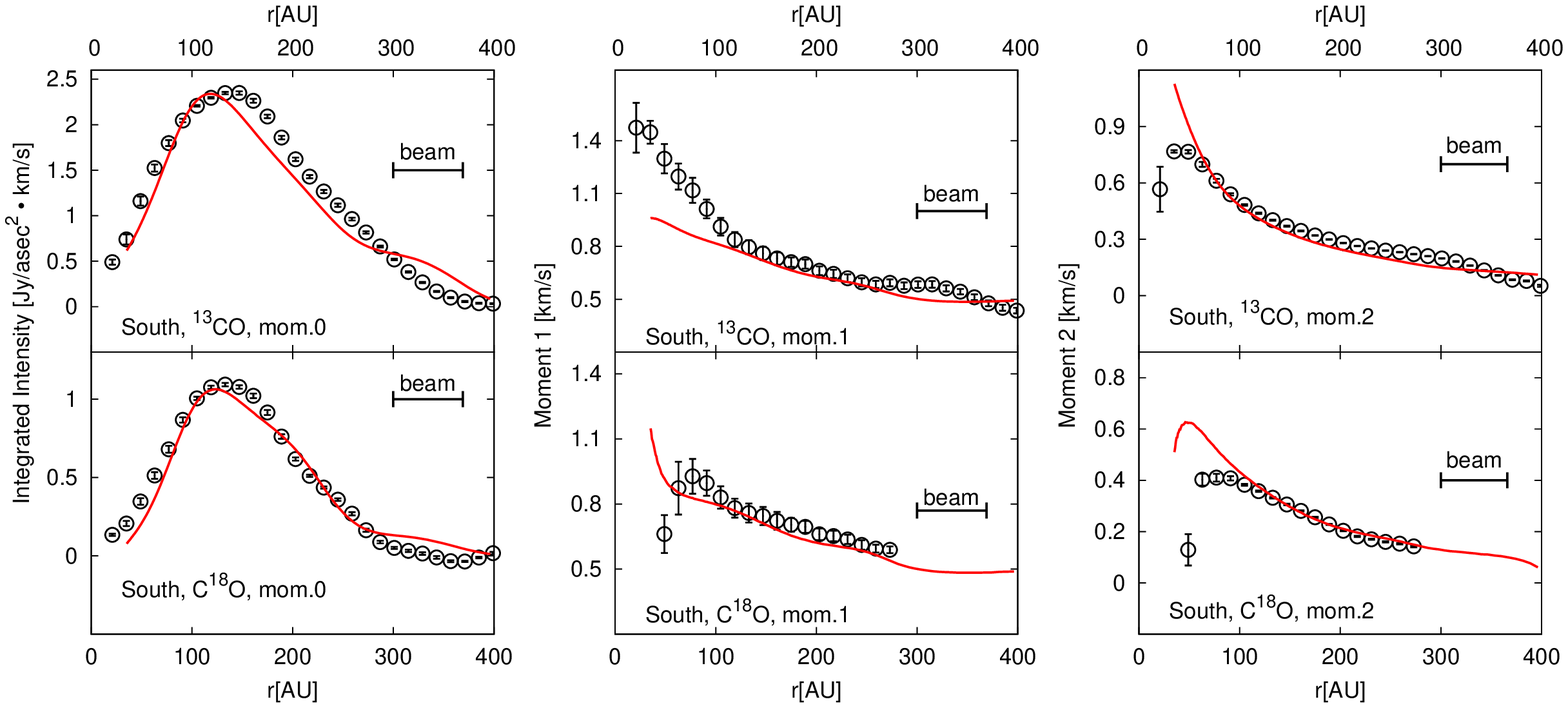} 
 \end{center}
 \caption{The same as figure \ref{fig:final_North} but for the south
 profiles.}
 \label{fig:final_South}
\end{figure*}

\begin{figure*}
 \begin{center}
  \FigureFile(15cm,12cm){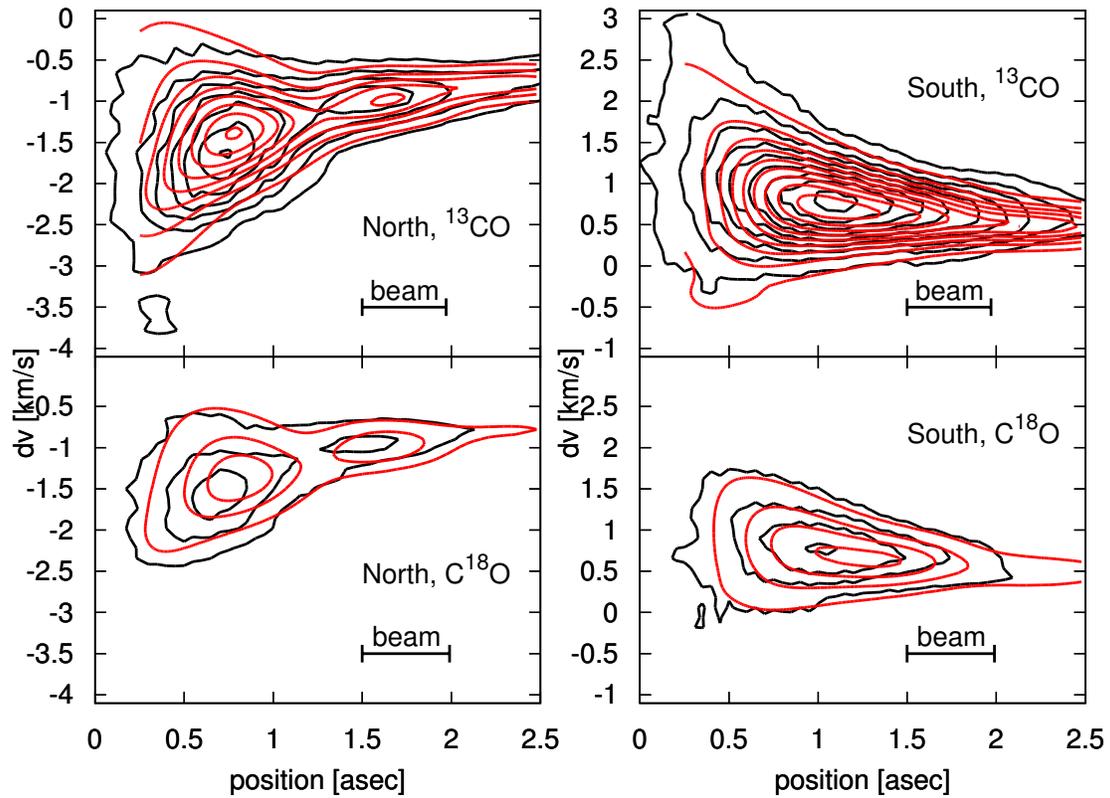} 
 \end{center}
 \caption{Comparison of P-V diagram between observations (black
 contours) and the model (red contours) for the north and south
 directions.  The left panels show the north direction and the right
 panels show south direction, while the top row shows the results for
 $^{13}$CO and the bottom row shows C$^{18}$O.  
 Contours start from 5-$\sigma$ level (0.135~Jy/asec$^2$
 for $^{13}$CO and 0.175~Jy/asec$^2$ for C$^{18}$O, respectively) with
 10-$\sigma$ step.}
 \label{fig:final_PVdiag}
\end{figure*}

The comparison of the radial profiles 
between the models and observations is given in
figures~\ref{fig:final_North} and \ref{fig:final_South} for the north
and south directions, respectively.
For these models, we show the comparison between the models and
observations of the moment 2 radial profiles averaged over 
$11^{\circ} \leq PA \leq 33^{\circ}$ (north) 
and $211^{\circ} \leq PA \leq 233^{\circ}$ (south) 
as well as 
the moments 0 and 1, which have been the main focus of the
modeling.  The radial profiles of all the moments show reasonable match
between observations and models.  
Finally, we show the comparison of P-V diagrams 
in the north and south directions for the reference models 
in figure \ref{fig:final_PVdiag}.  
The models and observations show reasonable
match not only in the radial profiles of the moment maps 
but also in the P-V diagram.

The radial profiles of the model moment 0 profiles exhibit 
some bumps and troughs while observations show more smooth
profiles.  
This is most prominent at around $\sim 170-200~\mathrm{AU}$ 
region of the north profiles where the dust emission is very bright.
These bumps are apparent structures due to the subtraction of bright
continuum emission, and we discuss about this further in section
\ref{sec:gasbump}.

\begin{table*}
 \begin{center}
 \begin{tabular}{ccc}
  ~ & North & South \\
  \hline
  \hline
  $\Sigma_0~[\mathrm{g~cm}$$^{-2}]$ & 0.845  & 0.2325 \\
  $f_{\rm in}$ & 1/8 & 1/8 \\
  $r_{\rm c}~[\mathrm{AU}]$ & 110 & 100  \\
  $r_{\rm out} [\mathrm{AU}]$ & 250 & 250 \\
  $f_{\rm out}$ & $10^{-2}$ & $10^{-1}$ \\
 \end{tabular}
 \end{center}
 \caption{Parameters of the reference models for gas distribution.}
 \label{table:finalvalues}
\end{table*}

\section{Discussions on the Reference Model}
\label{sec:discussion}

We have constructed the models of dust and gas distribution in the
protoplanetary disk around HD~142527.
In this section, we look at our reference models of dust and gas
distribution and discuss indications and caveats of our model.

\subsection{Summary of Gas and Dust Distribution}

\begin{figure*}
 \begin{center}
  \FigureFile(16cm,12cm){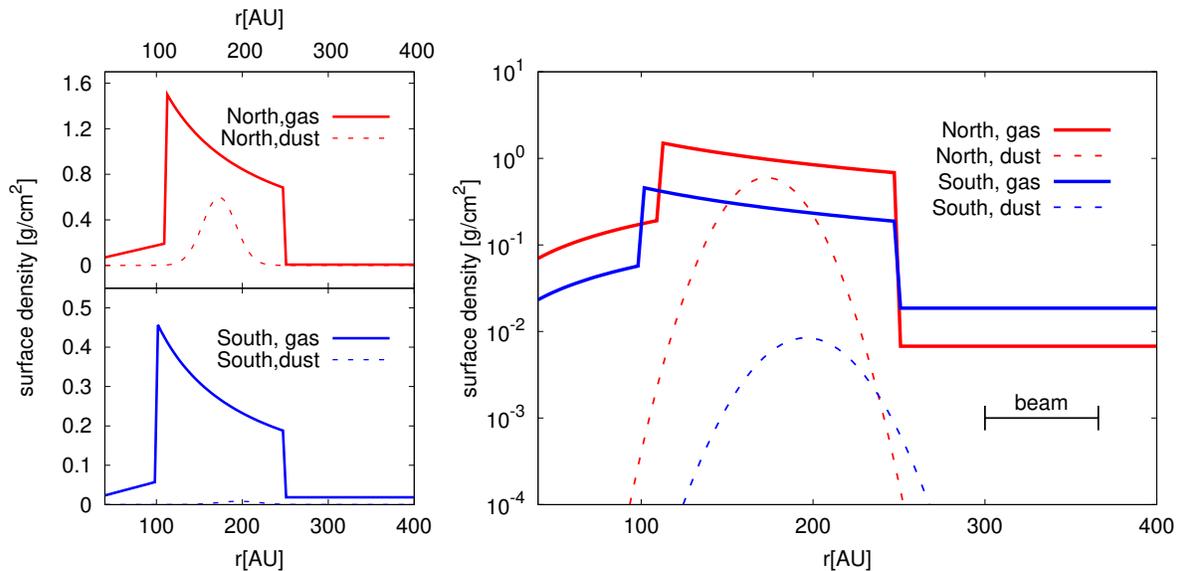} 
 \end{center}
 \caption{Surface density distribution of gas (solid line) 
 and dust (dashed line) in northern (red) and southern (blue) part of
 the disk.  The left panels show surface density in the linear scale
 (top panel for the north and the bottom panel for the south model),
 while the right panel shows the surface density in the log scale.}
 \label{fig:surfdens_final}
\end{figure*}

We have shown that the gas and dust are distributed very differently in
the disk. 
Figure~\ref{fig:surfdens_final} shows the surface density profiles of
gas and dust in the north and the south regions for the reference model. 
Dust distribution can be explained by the radial Gaussian ring-like
profile with the width of $w_d \sim 30$~AU (50~AU in FWHM)
for both the northern and southern regions. 
However, there is a factor of $\sim 70$ difference 
in surface density between the two regions  
(see equation \eqref{eq:dust_model} and
parameters listed in table \ref{table:dustvalues} for dust
distribution).  
Gas distribution is more or less axisymmetric and radially extended.   
The variation of surface density between the northern and
southern part is a factor of $\sim 3$ and 
the gas density cuts off at $r_{\rm c}\sim100$~AU, and 
$r_{\rm out}\sim250$~AU, but both inner and outer regions are not
completely devoid of gas (see equation \eqref{eq:finalmodel} and
parameters listed in table \ref{table:finalvalues} for gas
distribution). 

The gas-to-dust ratio is $\sim 30$ at the peak of the dust emission 
in the southern part of the disk, 
while it reaches $\sim 3$ in the northern peak.
Since dust emission is optically thick in the northern part of the disk,
the dust emission is not very sensitive to the amount of dust (see
appendix \ref{sec:app_dustdistribution}) so the amount of dust in the
northern part may be considered as a lower limit.  Therefore, the
gas-to-dust ratio may be even lower in the northern part of the disk.
The gas-to-dust ratio significantly lower than the 
canonical value (100 in the ISM) in the northern part 
indicates that the disk surface density (dust$+$gas) 
is not high enough to collapse into a protoplanet by gravitational
instability \citep{Fukagawa13}.

The radial distributions of dust and gas are fixed to Gaussian and
power-law (with modifications at inner and outer radii) functions,
respectively, in our modeling.  
The parameters that give a reasonable fit to observations are searched
manually.
Our model is one of the solutions but not unique, and we consider that
the numbers listed in table \ref{table:finalvalues} have at least a
factor of $\sim 2$ uncertainties, especially for $\Sigma_0$.
It is difficult to constrain the radial distributions more definately
since the spatial resolution is limited.
Still, we can argue that the gas distribution is more extended in the
radial direction than dust and the gas-to-dust ratio is lower in the
north than south.

At a glance at figure \ref{fig:surfdens_final}, the gas distribution
may be interpreted as ``radially extended bump'' and the dust particles
are strongly accumulated in this bump region.  
Although the locations of the 
peak of the continuum emission and gas moment 0 profiles are different
in the observed profiles (see figure \ref{fig:gasobs}),
our model indicates that the peak of dust surface density
profiles is within the region of gas surface density bump, which is
between $r \sim 100~\mathrm{AU}$ and $r \sim 250~\mathrm{AU}$.  The peak
of the gas emission at $\sim 100~\mathrm{AU}$ is partially due to the
fact that the gas temperature is higher at inner radii.  
Interestingly, the bump in gas seems to be weaker in the south region
than in the north.
Such distribution of dust and gas 
seems to be, at least qualitatively, consistent with the
picture of dust particles trapped in a pressure bump, or large-scale
vortex, in gas 
\citep{Pinilla12,Birnstiel13,Lyra13}.  
We also note that recent observations by ATCA at 34~GHz indicate that
large grains are concentrated in the northern part of the
disk~\citep{Casassus15}.
Vortices should be confined within a narrow radial range of 
the scale comparable with the disk thickness in this case.  
It is hinted that there might exist small-scale structures close to
the peak of dust distribution, as discussed at the end of 
section \ref{sec:gassummary}.
Future higher resolution observations may reveal the
small-scale structures within the bright dust ring.
Recently, \citet{Mittal14} proposed that the global $m=1$ mode 
might play an important role in producing radially extended lopsided
structures.   
One interesting prediction of this model is that dust grains with
different sizes accumulate at different locations within the disk.  
This may be addressed by higher resolution observations in
multi-wavelengths and modeling effort including several dust species.    
Another interesting mechanism for the formation of dust-rich regions is
the secular gravitational instability 
\citep{Ward00,Youdin11,Michikoshi12,Takahashi14}. 
\cite{Takahashi14} have done two-fluid analysis of gas and dust and shown
that (multiple) ring-like structure with low gas-to-dust mass ratio can
be created in self-gravitationally stable disks.  Since initial gas mass
should be relatively large in this case, significant gas dispersal
should have occurred after the ring formation.  
This could be addressed, for example, 
by investigating the detailed kinematics of gas (e.g.,
\citet{Pontoppidan11}).

\subsection{Optical Depth and the Total Amount of Dust and Gas}
\label{sec:optdepth}

\subsubsection{Dust Optical Depth and Dust Properties}
\label{sec:dust_tau}

\begin{figure*}
 \begin{center}
  \FigureFile(8cm,6cm){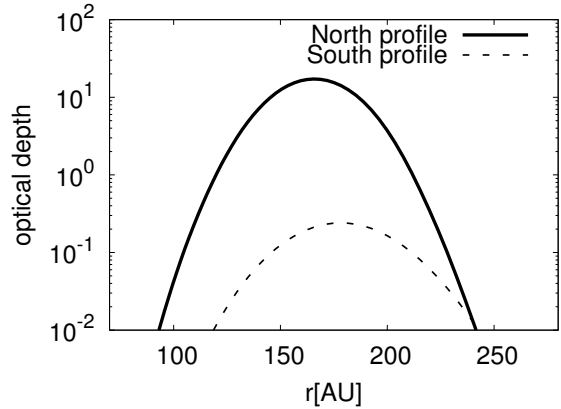} 
 \end{center}
 \caption{Optical depth profiles for continuum emission.  
 Solid line is for the north profile 
 and the dashed line is for the south profile.}
 \label{fig:tau_cont}
\end{figure*}

We have seen that the dust particles are strongly concentrated in the
north region.  
The radial profiles of the optical depth 
at the observed frequency ($\sim 340$~GHz)
along the line of sight is
shown in figure~\ref{fig:tau_cont}, and it is clear that,   
in the northern part, the optical depth of dust emission reaches 
$\gtrsim 10$.

\begin{figure*}
 \begin{center}
  \FigureFile(8cm,6cm){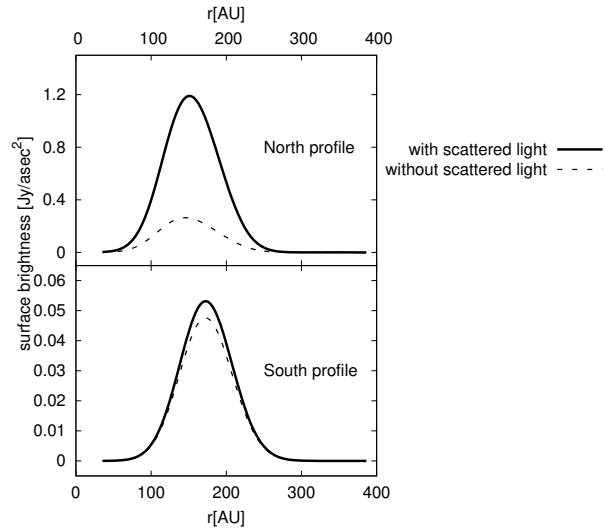} 
 \end{center}
 \caption{Radial profiles of continuum emission calculated with (solid
 lines) and without (dashed lines) dust scattering.  The left panel is
 for the north profile and the right panel is for the south profile.}
 \label{fig:compscat}
\end{figure*}

We have used the dust model with the maximum size of 1~mm. 
This is purely an assumption of this study, in order to estimate the
minimum amount of dust grains needed to explain the bright thermal
emission by using the dust model having the maximum 
(or at least large within the models that are considered to be
reasonable) opacity at sub-mm wavelengths.
In the case of dust model with $a_{\max}=1~\mathrm{mm}$,  
the dust continuum emission is dominated by scattered
light component if the disk is optically thick, which is the case for
the northern part, because the dust scattering coefficient is much
greater than the absorption coefficient (figure \ref{fig:dustmodel}).  
Figure \ref{fig:compscat} compares the radial profiles of dust
continuum emission between calculations with and without scattered light
component.  
We have artificially set the radiation energy
density $J_{\nu}=0$ in equation \eqref{eq:dustsource}
in the calculations to omit the scattered light component.  
In the northern part of the disk, 
the contribution from the scattered light component 
is a factor of $\sim 5$ larger than that from the thermal emission 
that has not experienced scattering.  
In the southern part, where the disk is optically thin, 
the contribution from the scattered light component is much smaller.  

There may be a variety of dust models that are able to reproduce
observations.  For example, the SED of HD~142527 can be reproduced by 
using irregularly-shaped micron-sized dust particles \citep{Verhoeff11}.   
We have checked that the slope of SED in sub-mm range is
consistent with observations in our model.  
Therefore, it is difficult to discriminate 
the dust size contributing most to sub-mm emission 
from current observations thus far.  
The scattering coefficient $\kappa_s$ of micron-sized dust particles 
at sub-mm range is expected to be much smaller than  
that used in our work.  
If the scattered light component can be observed exclusively, it is
possible to discriminate the dust size.  
The polarization of dust continuum emission, for example,  
can be a good tracer of the dust size~\citep{Kataoka15}.

\subsubsection{Gas Optical Depth}
\label{sec:gas_tau}

\begin{figure*}
 \begin{center}
  \FigureFile(8cm,6cm){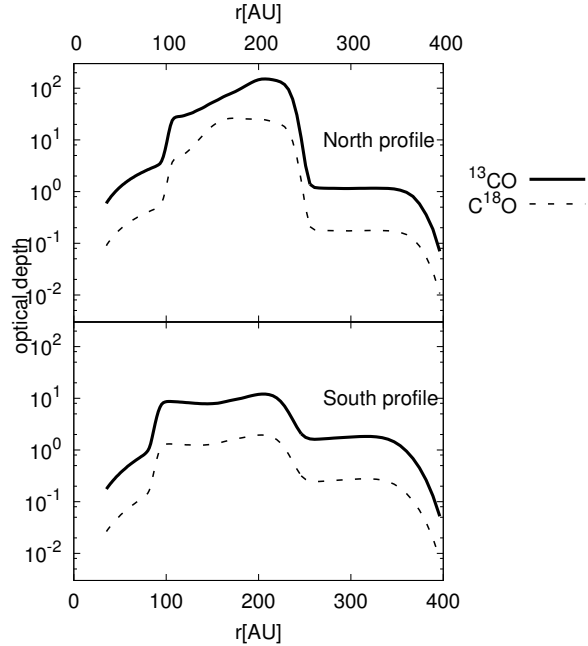} 
 \end{center}
 \caption{The line-of-sight optical depth of $^{13}$CO (solid line) and
 C$^{18}$O (dashed line).  Top panel shows the optical depth of the
 northern part of the disk and the bottom panel shows that of the
 southern part.}
 \label{fig:tau_gas}
\end{figure*}

\begin{figure*}
 \begin{center}
  \FigureFile(16cm,12cm){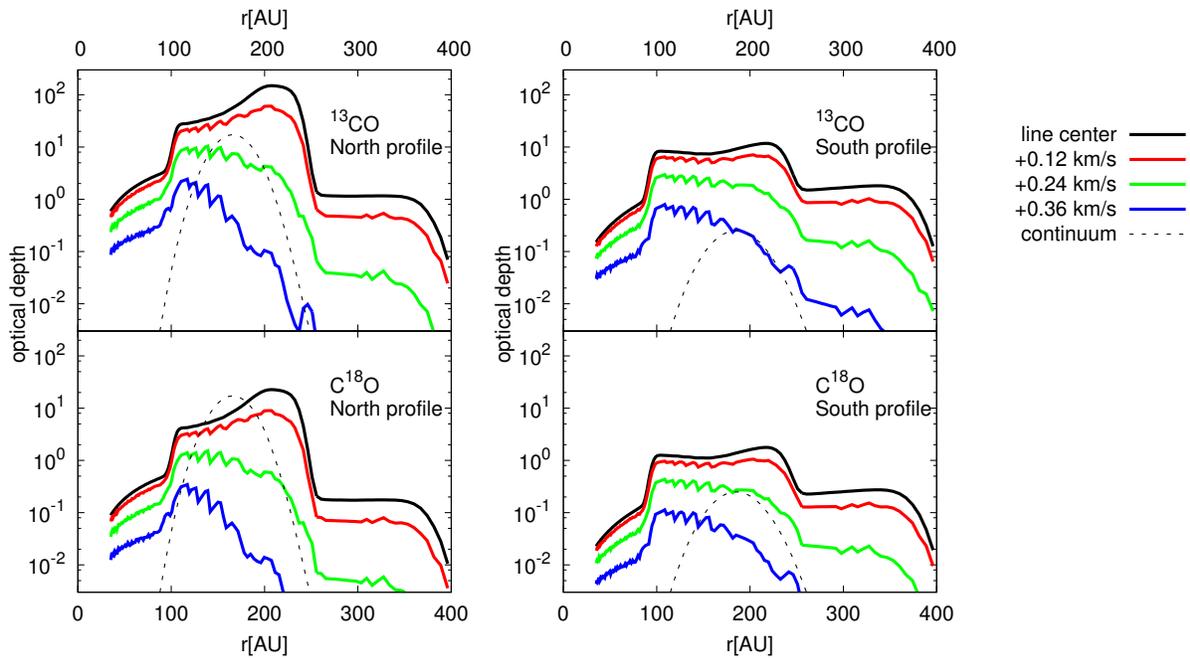} 
 \end{center}
 \caption{
 The line-of-sight optical depth of the north profile (left) and
 south profile (right).  The upper panels show the optical depth of 
 $^{13}$CO and the lower panels show that of C$^{18}$O.  
 In each panel, the line-of-sight optical depth of the line center (the
 maximam optical depth of all the channels) is shown by black solid
 lines and that of the channels 
 offset from the line center by 0.12~km/s, 0.24~km/s, 0.36~km/s is shown
 in red, green, and blue lines respectively.  The optical depth of the
 continuum emission (black dashed lines) is subtracted for all of the
 gas optical depth.
}
 \label{fig:tau_gas_offset}
\end{figure*}

We have used $^{13}$CO $J=3-2$ and C$^{18}$O $J=3-2$ emission 
to determine the gas structure.
Figure \ref{fig:tau_gas} shows the radial profiles of the 
maximum optical depth of these lines (gas$+$dust optical depth, see
also figure \ref{fig:tau_gas_offset}) for the reference disk models.  
It is shown that the gas is optically thick at the peak of the lines.  
The maximum optical depth reaches an order of unity
even for the south model of C$^{18}$O.

Despite the fact that the lines are mostly optically thick, 
we have seen that the integrated intensity of gas emission 
becomes brighter as we increase the
gas surface density (figures \ref{fig:varsig0_North} 
and \ref{fig:varsig0_South}).  This is because gas emission is not
entirely optically thick for all the velocity channels.  
When the amount of gas is
increased, gas emissivity is increased and therefore, there are more
velocity channels that contribute to the gas brightness at each given
spatial location.  
Figure \ref{fig:tau_gas_offset} shows the radial profiles of the optical
depth of the line wings.  
We show the optical depth at the velocity 
channels that are different from those giving the maximum optical depth
(line center) by 0.12~km/s, 0.24~km/s, and 0.36~km/s.  For all of the
calculations shown in figure \ref{fig:tau_gas_offset}, 
the optical depth of the continuum emission (which is derived using the
first term of equation \eqref{eq:chi_ul}) is
subtracted from gas+dust optical depth (which is derived using all the
terms using the right hand side of \eqref{eq:chi_ul}) 
to show the gas-only optical depth\footnote{
In other words, gas-only optical depth is
calculated by taking into account only the second term of the right hand
side of equation \eqref{eq:chi_ul}}.
The line emission becomes
optically thin at $0.2-0.3$~km/s away from the line center. 
Since the values of moment 2 shown in figure \ref{fig:final_North} and
\ref{fig:final_South} are $\sim 0.3$~km/s at $r\sim 150$~AU, 
we consider that the line emission is optically thin 
when the velocity deviates from the line center by (only) half width of
the line, which is close to the value of moment 2.
Therefore, line wings can be used to trace the gas surface
density even in the case that the emission at line center is optically 
thick.
We therefore consider that we can reasonably 
constrain the amount of gas even when it is optically thick.  
However, we yet see that $^{13}$CO is more insensitive to the change of
the amount of gas compared to C$^{18}$O, which is more optically thin.
Overall, we expect that the amount of gas has an uncertainty of a factor
of $\sim 2-3$.  For example, figure \ref{fig:varsig0_North} indicates the
models with $\Sigma_0=2.325~\mathrm{g~cm}^{-2}$ and 
$\Sigma_0=0.845~\mathrm{g~cm}^{-2}$ 
both explain the radial profiles of moment 0 in the north direction 
at $r = 150-200~\mathrm{AU}$ region reasonably well.  
To better constrain the amount of gas, we may need observations of lines with
lower optical depth by, for example, using rarer isotopologues or
using lower transition lines.  
The values of isotope ratio and the dust opacity are also 
important in accurately determining the amount of gas.
Since there is fair amount of dust, the line emission is
affected by the dust absorption, especially in the northern part of the
disk.
The scattering of gas lines by dust particles, which is not
included in our model,  
can also be effective in determing the observed spatio-kinematic
patterns.

\subsubsection{Total Amount of Gas and Dust}
\label{sec:totalamount}

We now discuss the total amount of dust and gas indicated from our
modeling.  We have two surface density distribution models,
namely the north and south model, for each dust and gas.

For dust mass, we obtain 
$M_{\rm dust}=3.5\times 10^{-3} \msun$ for the north profile and 
$M_{\rm dust} = 7 \times 10^{-5} \msun$ for the south profile
when integrated over the entire disk.  
\citet{Verhoeff11} obtained the dust mass of $1\times 10^{-3} \msun$ 
from their SED modeling.  Considering that the 
bright arc of dust emission extends about $\sim 1/3$ 
of the whole azimuth, the value of dust mass we have obtained 
is similar to their results, despite the difference of grain properties.   
For gas mass, we obtain 
$M_{\rm gas} = 1.8 \times 10^{-2} \msun$ and 
$M_{\rm gas} = 5.7 \times 10^{-3} \msun$ 
for the reference north and south models, respectively, 
when integrated over the disk 
within 400~AU.\footnote{Inside 400~AU, the gas mass
is dominated by the gas residing within $r\lesssim r_{\rm out}$ although
we have assumed that constant surface density in the radial direction
for outer teneous gas.}

Although the total gas mass could have a factor of $\sim 2-3$ 
uncertainty as noted before, 
the overall gas-to-dust mass ratio may be $\lesssim 10-30$, 
which is smaller than the interstellar value of 100.
The derived gas-to-dust mass ratio is likely to be an upper limit.  
Since the north region is optically thick, the dust emission is not very
sensitive to the increase in the amount of dust 
(see appendix \ref{sec:app_dustdistribution}).  
We should also remember that  
the dust particles have relatively large opacity (and therefore
emissivity) in sub-mm range ($a_{\max}=1~\mathrm{mm}$).
If smaller grains are more abundant, 
we expect that the disk midplane, where dust
emission originates, is colder while the disk surface, where optically
thick gas emission originates, is warmer \citep{Inoue09}.  
Consequently, even if small grains are more abundant, 
we expect brighter emission in gas
while fainter in dust, resulting in smaller gas-to-dust ratio.
We therefore propose that the dispersal of
gas, or at least the depletion of CO gas, may have already taken
place in the disk of HD~142527.  
However, it should be noted that detailed modeling with different dust
grain properties is necessary to confirm this.  Grain size and shape can
strongly affect the observed emission properties (e.g., \cite{Min03}).  

\citet{Perez14} derived the total gas mass of
0.1~$\msun$ from their modeling effort.  
They used the model where the temperature is assumed to be 
50~K at 50~AU with $r^{-0.3}$ profile and the Doppler $b$ parameter is
assumed to be 50~m~s$^{-1}$.
Their radial temperature profile is flat and the temperature is
smaller than our model by a factor of $\sim 1.7$ 
at $r=100~\mathrm{AU}$.
We have calculated how the assumptions of 
the temperature and the line width affect the results by setting
those parameters to be the same as \citet{Perez14} but the density is
given by our reference model 
(equation \eqref{eq:finalmodel} and table \ref{table:finalvalues}).
The values of $^{13}$CO $J=3-2$ moment 0 is smaller 
by a factor of $\sim 2$ at $r \sim 100$~AU with this model.  
The impact of line width on moment 0 profile comes from
the fact that there are less velocity channels 
that contribute to integrated intensity at each spatial position 
(see also the discussion in section \ref{sec:gas_tau}).  
To compensate this by varying the amount of gas (parameter $\Sigma_0$ in
our model), we need to increase the value of $\Sigma_0$ by 
a factor of $\sim 5-10$, as seen in figure~\ref{fig:varsig0_North}.  
Consequently, we consider that the difference of 
the total amount of gas comes from the different assumption of the
temperature and the line width.
The temperature profile used by \citet{Perez14} comes from 
$^{12}$CO $J=2-1$ observations, which may be smoothed by the relatively
large beam.
High resolution gas observations will play a
decisive role in determining the distribution and total mass of gas
component more accurately.

\subsection{Gas Bump Structures in Model}
\label{sec:gasbump}

The north radial profiles of the moment 0 of the reference model 
is bumpy at $r\sim 170-200$~AU, where dust emission is the brightest.
As discussed in section \ref{sec:varsig0}, 
these bumps and troughs in gas emission 
arise because the continuum emission is very bright and comes from
relatively narrow radial range.
Since the line emission is partially hidden by dust, there is a
significant effect on line emission when dust continuum is subtracted.

There are at least two possible ways to obtain smoother moment 0
profiles. 
One possibility is to increase the amount of gas at the location 
of the peak of dust emission locally.  
In other words, there may be more detailed, 
small-scale (several tens of AU scale) structures 
in the radial direction than we have considered.
As seen in figure \ref{fig:varsig0_North}, 
a factor of $\sim 3$ larger gas surface density locally 
at the peak of dust distribution may explain the observed 
values of moment 0 at this region.
Since the dust emission is barely resolved in the radial direction, it
may be difficult to see such small-scale variation of gas distribution.  
Higher angular resolution observations of gas is necessary to confirm
this possibility.

The other possibility is to consider more optically thin dust particles.
Since the dust scattering opacity is $\sim 10$
times larger than the absorption opacity in our dust model, 
the total extinction coefficient of dust is dominated 
by the scattering opacity and is comparable with gas opacity 
(figure \ref{fig:tau_gas_offset}).  
Therefore, if the dust scattering coefficient is much smaller than
considered in this paper, the gas emission can be brighter so the
effects of continuum subtraction is more insignificant.  
Smaller dust particles may be one solution, but in this case, it is
necessary to have larger amount of dust because the absorption
coefficient (and therefore dust emissivity) is also small.

A large (effective) emissivity and small scattering coefficient may be
obtained simultaneously if we consider dust sedimentation.    
In our framework, the gas-to-dust ratio is constant in the vertical
direction so large ($\sim$mm in size) dust particles reside even in the
upper layer of the disk.  
However, such large particles 
may be sedimented to the disk midplane.  
The surface of the CO emission resides in the upper layer of the
disk, and therefore there may only be small dust particles having small
scattering coefficient around the gas surface.  
The gas emission is not hidden by dust very much in this case 
and therefore the effects of the subtraction of the continuum emission 
is less significant.  
More sophisticated models that take into account 
the dust particle motion is necessary to verify this possibility.  
Also, the optical properties of the dust particles and its impact on
observations should be carefully investigated.
Detailed modeling in tandem with scattered light observations in near
infrared (NIR; see also section \ref{sec:outer_NIR}), 
which is sensitive to the small dust particles in the
upper layer of the disk, may be a key to verify this possibility.

\subsection{Discrepancy of Moment 1 at Inner Radii}

In the moment 1 profiles of the south gas models, 
we see that the reference model shows slower speed than observed  
at $r<100~\mathrm{AU}$ for $^{13}$CO
but faster at $r<50~\mathrm{AU}$ for C$^{18}$O, as shown 
in figure \ref{fig:final_South}.  
We have checked that this discrepancy is also present when the surface
density profiles for the remaining gas inside the cavity is taken to be
constant so the details of the radial profile of gas surface density
do not alter the results.  
C$^{18}$O data may be affected by the lower signal-to-noise ratio at high
velocity channels, but the detection of $^{13}$CO at
$r\sim50-100~\mathrm{AU}$ region is robust.

It is natural that the observed velocity should become smaller than
Keplerian as one goes to inner radii, 
where the disk is barely spatially resolved in 
channel maps and the signal is weak.  
The gradient of line-of-sight velocity within the beam is large at inner
radii and therefore emission at many velocity channels are averaged.
Weak high velocity component may be discarded 
when the sensitivity is limited.
These effects should properly be incorporated in our modeling 
since we convolve each channel map image by the Gaussian beam before
calculating the model moment maps (see section \ref{sec:method_gas}).
Therefore, the discrepancy between the
observations and models might indicate that, in
the southern region, the gas velocity is significantly faster
than Kepler velocity at least for $^{13}$CO at inner radii.  
\citet{Rosenfeld14} discussed that there may be a fast radial flow in
the inner region of the disk of HD~142527 
based on $^{12}$CO and HCO$^{+}$ data taken by ALMA Cycle 0.
\citet{Marino15} suggested that the inner disk is significantly inclined
relative to the outer disk based on the modeling of scattered light in
the H-band.  
The discrepancy between the reference model and
observations of $^{13}$CO for the southern region may also indicate the
existence of 
such dramatic change of the inner disk structure.  
Observations with better spatial resolution 
are essential to clarify this point.

\subsection{Outer Tenuous Gas and Scattered Light in Near Infrared}
\label{sec:outer_NIR}

We have seen that there should be some amount of tenuous gas at
$r>250~\mathrm{AU}$.  The surface density of the tenuous gas is
estimated to be 
$\sim 7\times 10^{-3}-2\times10^{-2}~\mathrm{g/cm^2}$.
It is known that the HD~142527 disk shows 
an extended ($\sim 300~\mathrm{AU}$ scale) 
scattered light emission in NIR observations  
\citep{Fukagawa06,Casassus12,Canovas13,Rodigas14}.
Especially, the large-scale spiral structure appears from the south of
the disk and extends towards the west.

The extended scattered light emission in NIR may be connected to the
existence of tenuous gas at outer radii.  In NIR direct
imaging observations, scattered light by small 
(typically, $\sim 1~\mu\mathrm{m}$ in size) dust grains at the disk surface 
is observed.  
Since such particles are well coupled with gas, 
we expect that small grains are distributed as gas is.  
The amount of tenuous gas 
($\sim 0.02~\mathrm{g~cm}^{-2}$ at $r \sim 300~\mathrm{AU}$ in the south
model) is probably enough to make the disk optically thick 
in NIR range,  
where dust opacity may be several tens of cm$^2$ per unit gram of gas  
in the case of $a_{\max}\sim 1\mu$m if gas-to-dust ratio is 100 
in the outer $r>250$~AU region (e.g., \cite{DAlessio01,Aikawa06}). 
Considering the detection limit of dust continuum, we expect that such
(small) grains contained in the outer tenuous gas component are neither
observable nor massive enough to contribute to the gas-to-dust ratio
within 400 AU (see section \ref{sec:totalamount}).
To further investigate the distribution of small grains, we need
simultaneous modeling of scattered light and mm-emission, which is
beyond the scope of this paper.

\section{Conclusion}
\label{sec:summary}

\begin{figure*}
 \begin{center}
  \FigureFile(10cm,10cm){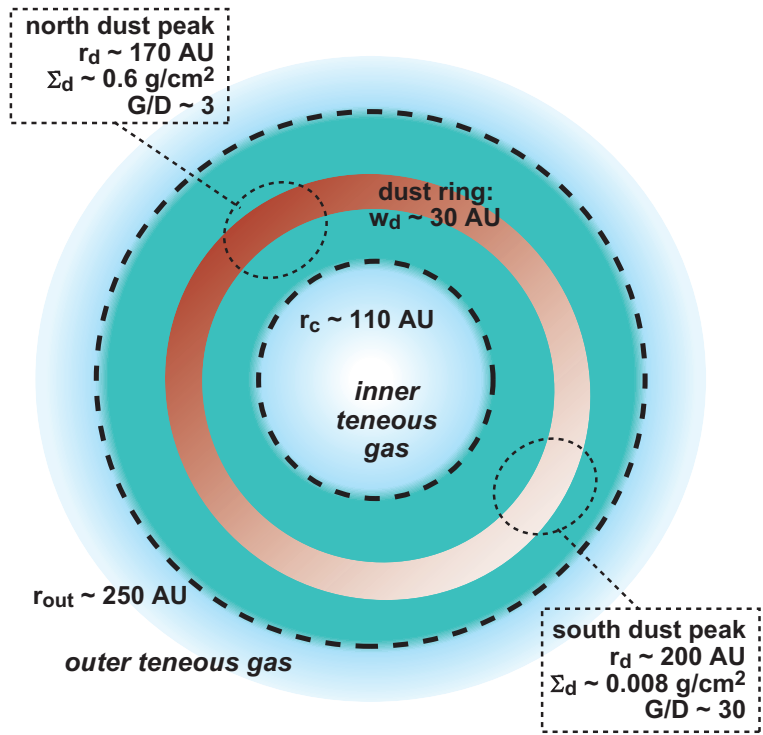} 
 \end{center}
 \caption{Schematic picture of the disk around HD~142527.  Green colors
 indicate gas distribution and brown indicate dust distribution.  
 The
 dust ring with the width of $w_d \sim 30~\mathrm{AU}$ is located at 
 $r \sim 170-200~\mathrm{AU}$.  The ring surface density is differ by a
 factor of $\sim 70$ in the northern and the southern part.  The gas
 surface density is more radially extended.  The gas distribution shows
 a bump-like structure from $\sim 100~\mathrm{AU}$ 
 to $\sim 250~\mathrm{AU}$ but there is remaining, tenuous gas component
 both inner and outer radii.}
 \label{fig:cartoon}
\end{figure*}

We have presented the models for gas and dust distribution for the disk
around HD~142527 by using a series of axisymmetric models and by
comparing the radial profiles of dust and gas emission.  
The schematic picture of the gas and dust distribution 
is shown in figure~\ref{fig:cartoon}.
Below, we list the main conclusions of our work.

\begin{enumerate}
 \item The dust particles are confined in a ring 
       whose surface density profile is described by a Gaussian
       function in the radial directions with 
       the width of $w_d \sim 30~\mathrm{AU}$ for both the north and south
       regions.  The location of the ring
       is slightly different between the north 
       ($r_d \sim 170~\mathrm{AU}$) 
       and south profiles ($r_d \sim 200~\mathrm{AU}$).
       To account for the
       strong azimuthal asymmetry of the continuum emission, 
       the surface density of the dust ring is 
       different by a factor of $\sim 70$ between the
       northern and southern part of the ring.  
       Note that these results are based on the dust model with
       the maximum size of 1~mm, where the opacity at sub-mm wavelengths
       is large and therefore, we expect that the dust mass derived in
       our work is close to the lower limit.
 \item The gas distribution is azimuthally more symmetric than dust
       distribution.  
       The amount of gas in the northern region is indicated to be 
       $\sim 3$ times as large as that in the southern region.  
 \item The gas radial surface density distribution is more radially
       extended than dust.  The radial profiles of the moment maps of
       $^{13}$CO~$J=3-2$ and C$^{18}$O~$J=3-2$ emission 
       can be described with the power-law profile 
       between $\sim 100~\mathrm{AU}$ and $\sim 250~\mathrm{AU}$, along
       with the remaining gas inside the cavity 
       and with tenuous outer gas at $r>250~\mathrm{AU}$.
       Small-scale ($\sim 50~\mathrm{AU}$-scale) variation in the radial
       directions of gas distribution with a factor of $\sim 3$ is also
       indicated around the peak of dust distribution. 
 \item Gas-to-dust ratio varies significantly within the disk.   
       At the peak of the dust distribution, it is $\sim 3$ 
       in the northern part of the disk while 
       $\sim 30$ in the south.
       The gas-to-dust ratio over the whole disk is
       expected to be $\sim 10-30$ within 400~AU, 
       indicating that gas depletion has
       already occurred in this system.
\end{enumerate}

We have used a series of axisymmetric disk models to do systematic
parameter search while keeping the problem tractable.  
Since the CO isotopes used in this
observation have turned out to be optically thick, observations of rarer 
isotopes or lower transitions are necessary to better constrain the
amount of gas. 
More elaborate dust models will also be required
to fully constrain the disk parameters and to account for observations 
at other wavelengths.

\bigskip

We thank Sebastian Perez for useful discussions.  
We also thank an anonymous referee for careful reading and useful
suggestions to improve the paper.
This paper makes use of the following ALMA data: 
ADS/JAO.ALMA\#2011.0.00318.S. 
ALMA is a partnership of ESO (representing its member states), 
NSF (USA), and NINS (Japan), 
together with NRC (Canada) and NSC and ASIAA (Taiwan), 
in cooperation with the Republic of Chile.
The Joint ALMA Observatory is operated by ESO, AUI/NRAO and NAOJ.
This work is partially supported by 
JSPS KAKENHI Grant Numbers 23103004, 26800106 and 26400224.

\appendix

\section{Details on $^{13}$CO and C$^{18}$O results}

\begin{figure*}
 \begin{center}
  \FigureFile(17cm,22cm){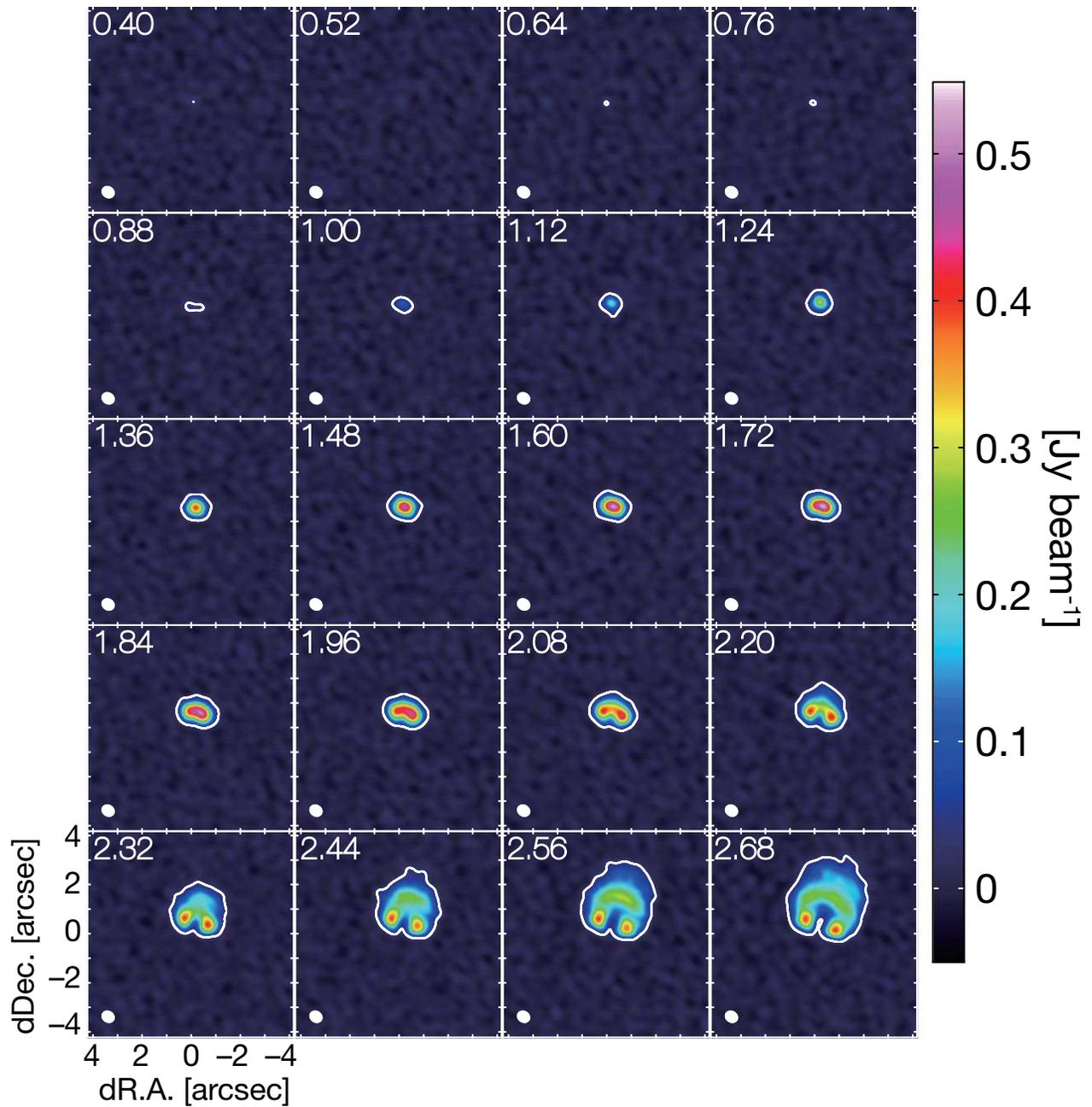} 
 \end{center}
\caption{Channel maps of $^{13}$CO ($J=3-2$). The LSR velocity is shown 
in the top left corner of each panel, and 
the synthesized beam, $0\farcs50 \times 0\farcs42$ with 
the major axis $PA=57.4^{\circ}$, is indicated by the ellipse 
in the bottom left corner of each panel. 
The white contours are the $5\sigma$ level, or 32 mJy beam$^{-1}$.}
\label{fig:ch-13co1}
\end{figure*}

\begin{figure*}
 \begin{center}
  \FigureFile(17cm,22cm){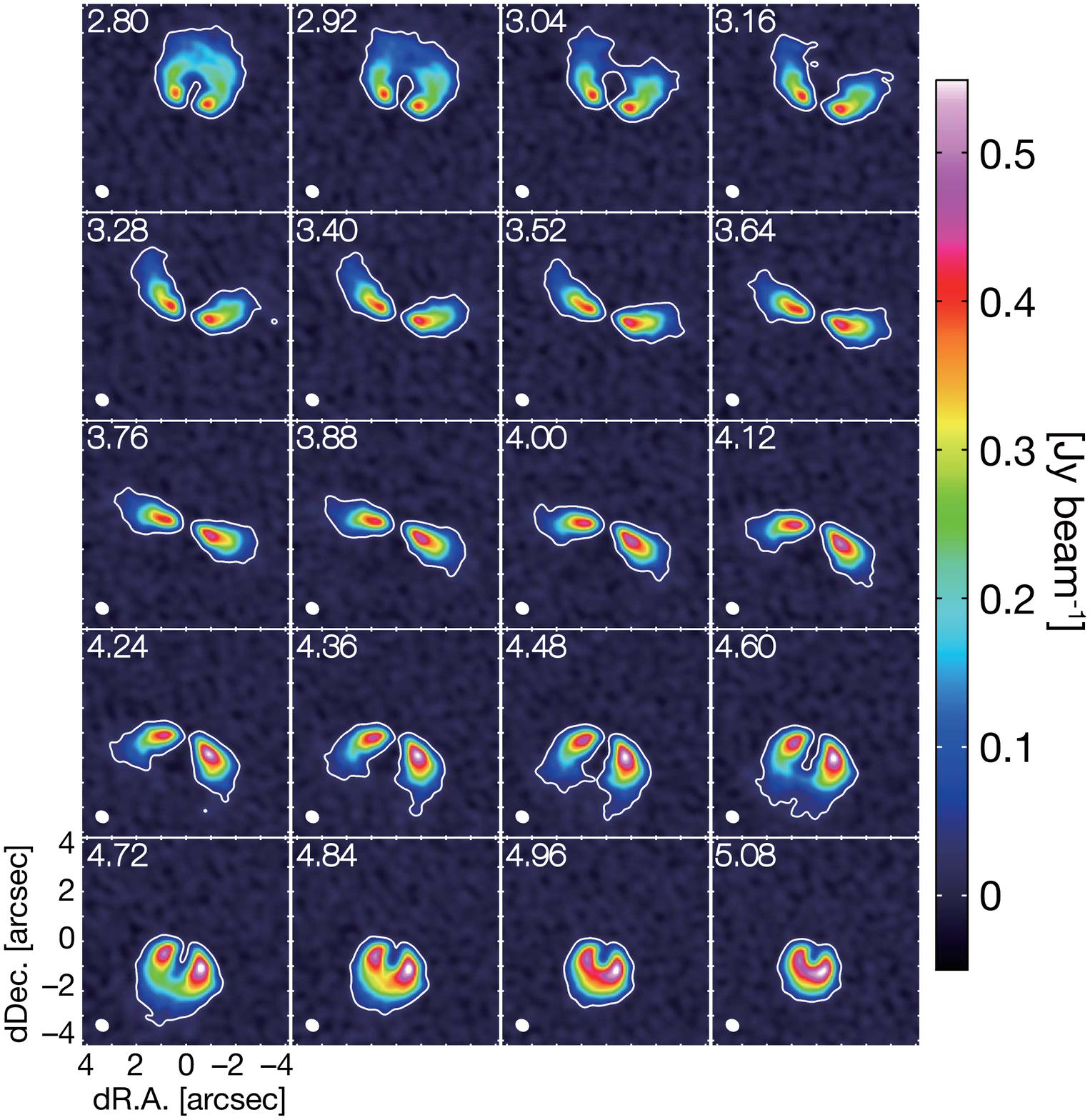} 
 \end{center}
\caption{Channel maps of $^{13}$CO ($J=3-2$), continued from 
figure \ref{fig:ch-13co1}.}\label{fig:ch-13co2}
\end{figure*}

\begin{figure*}
 \begin{center}
  \FigureFile(17cm,22cm){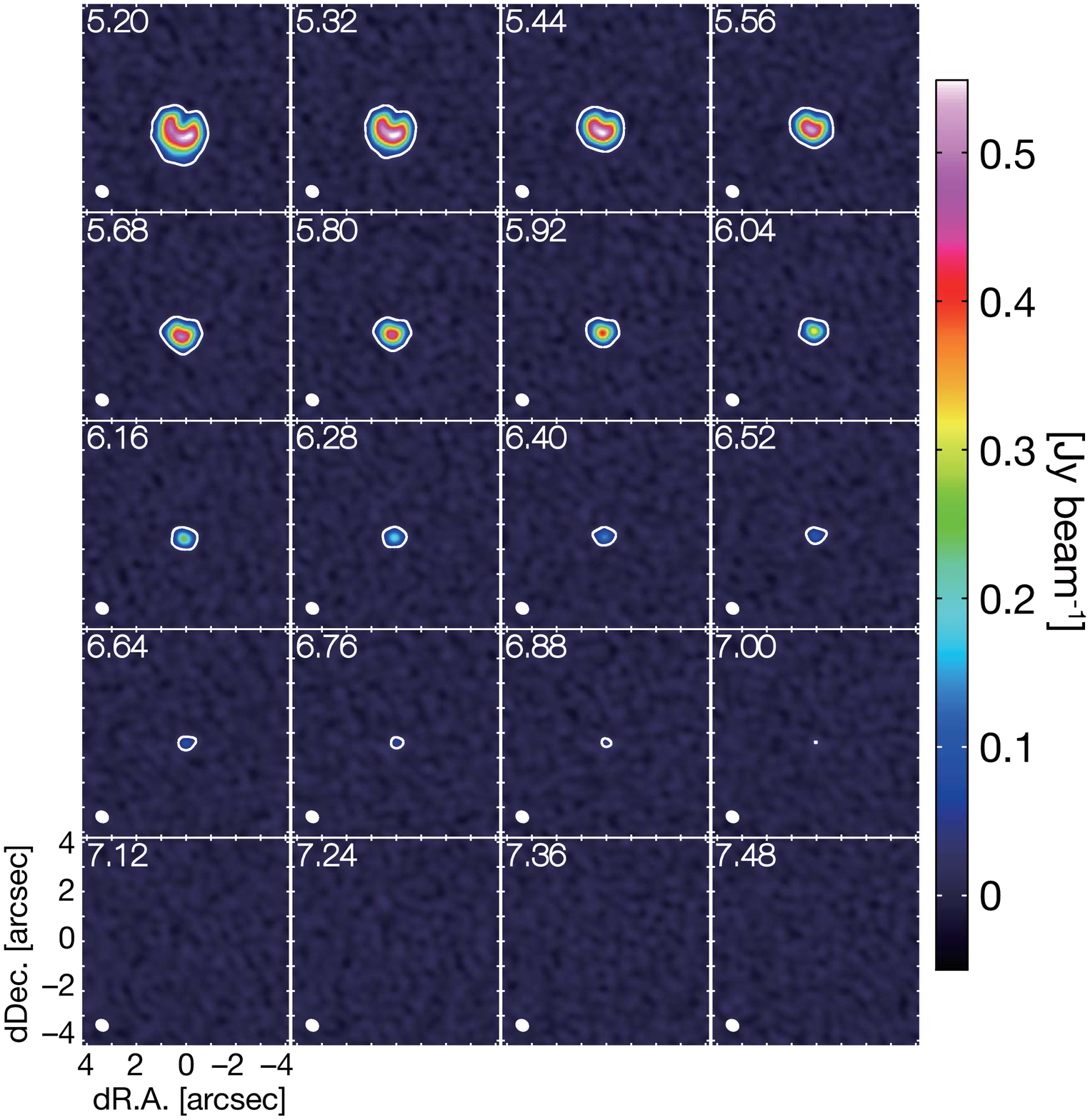} 
 \end{center}
\caption{Channel maps of $^{13}$CO ($J=3-2$), continued from 
figure \ref{fig:ch-13co2}.}\label{fig:ch-13co3}
\end{figure*}

\begin{figure*}
 \begin{center}
  \FigureFile(17cm,22cm){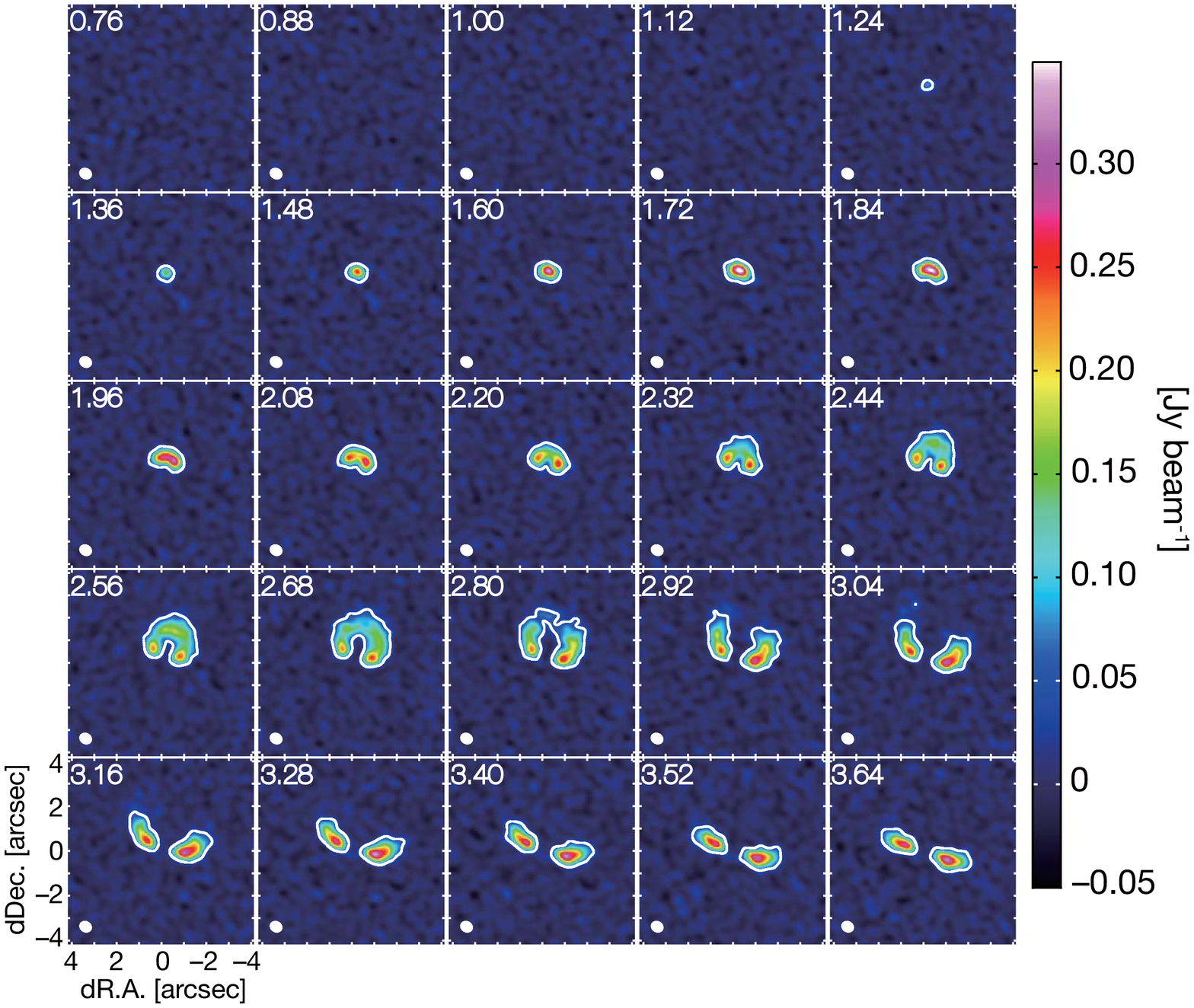} 
 \end{center}
\caption{Channel maps of C$^{18}$O ($J=3-2$). The LSR velocity is shown 
in the top left corner of each panel, and 
the synthesized beam, $0\farcs50 \times 0\farcs42$ with 
the major axis $PA=60.6^{\circ}$, is indicated by the ellipse 
in the bottom left corner of each panel. 
The white contours are the $5\sigma$ level, or 41.5 mJy beam$^{-1}$.}
\label{fig:ch-c18o1}
\end{figure*}

\begin{figure*}
 \begin{center}
  \FigureFile(17cm,22cm){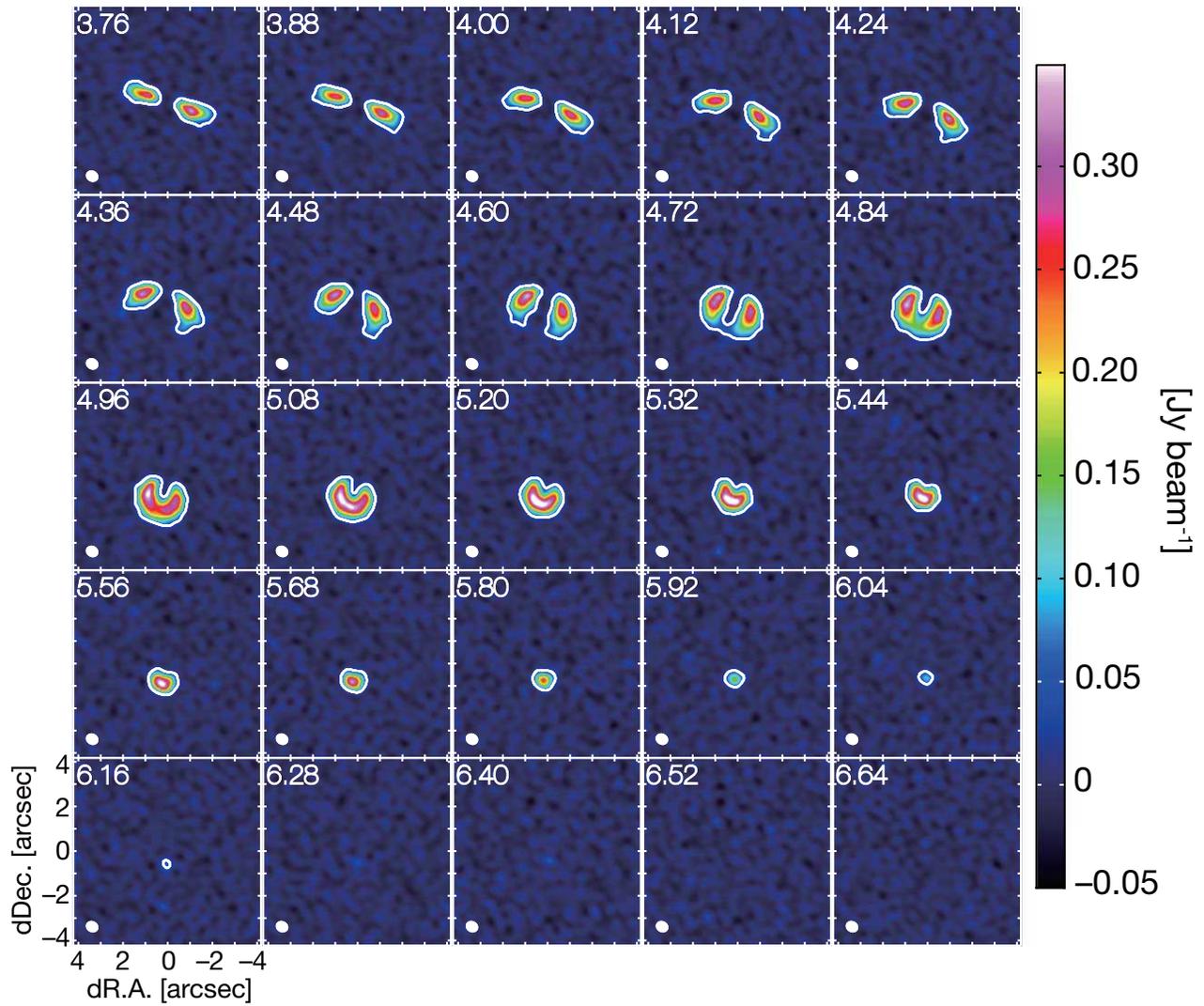} 
 \end{center}
\caption{Channel maps of C$^{18}$O ($J=3-2$), continued from 
figure \ref{fig:ch-c18o1}.}\label{fig:ch-c18o2}
\end{figure*}

Figures \ref{fig:ch-13co1}, \ref{fig:ch-13co2} and \ref{fig:ch-13co3} 
show the channel maps of $^{13}$CO from which the moment maps 
shown in figure \ref{fig:13co_mom} are created. 
Figures \ref{fig:ch-c18o1} and \ref{fig:ch-c18o2} 
show the channel maps of C$^{18}$O from which the moment maps 
shown in figure \ref{fig:c18o_mom} are created. 

\begin{figure*}
 \begin{center}
  \FigureFile(15cm,12cm){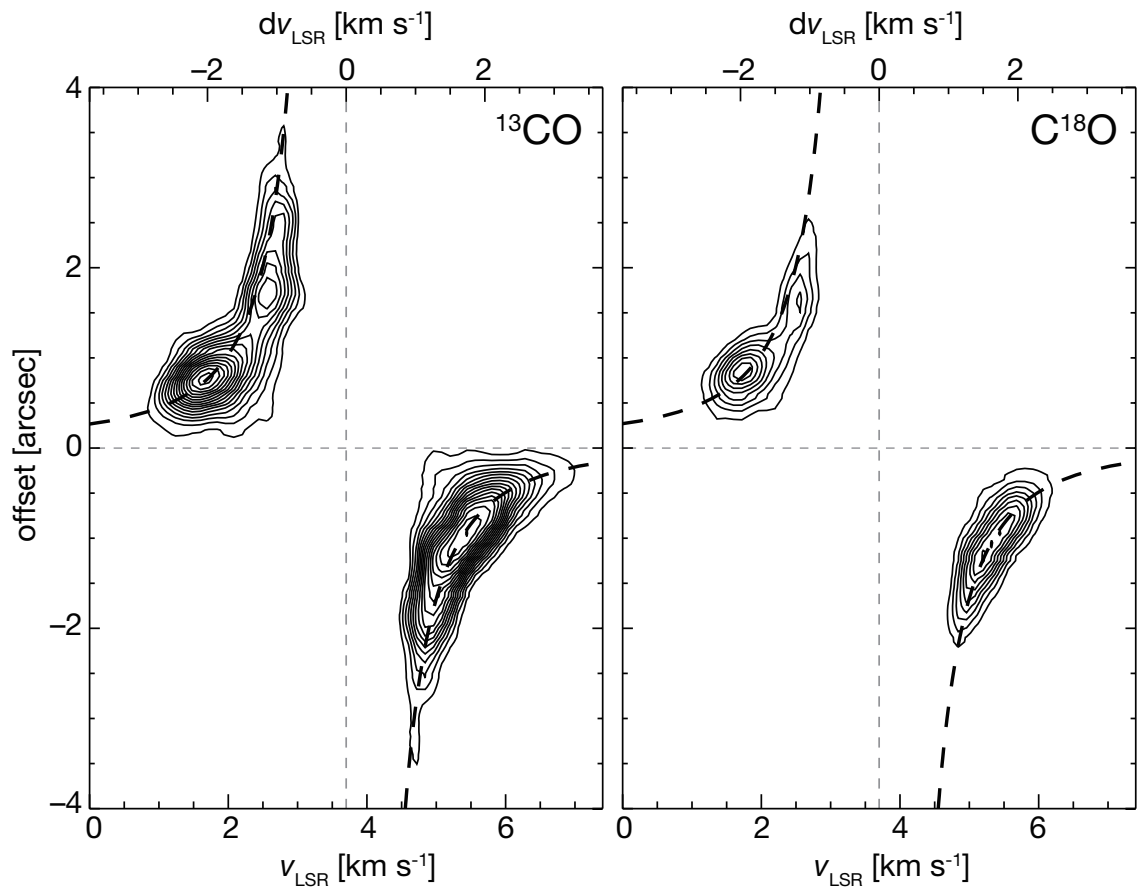} 
 \end{center}
\caption{The PV diagrams along the major axis ($PA=341^{\circ}$) in 
$^{13}$CO and C$^{18}$O $(J=3-2)$ lines. The contour spacing is $5\sigma$, 
starting at the $5\sigma$ level. The dashed curves in each panel 
indicate the Keplerian rotation when $M=2.2 M_{\odot}$ and the 
inclination angle $i=27^{\circ}$.}\label{fig:pv_major}
\end{figure*}

As described in section \ref{results-co},
a constant (systemic) velocity of 3.7~km~s$^{-1}$ is 
found along $PA = 71^\circ \pm 2^\circ$, which can be 
regarded as the direction of the minor axis of the system. 
The position-velocity (P-V) diagram along the 
major axis ($PA = -19^\circ$; figure~\ref{fig:pv_major})
was fitted by circular Keplerian motion for the 
emission detected above $5\sigma$ (figure~\ref{fig:pv_major}). 
The systemic velocity, position of the center of mass (the central 
star), and inclination relative to an observer are set as free 
parameters, whereas the central stellar mass is fixed in the range 
of $2.2 \pm 0.3$~$M_{\odot}$ (Verhoeff et al., 2011). 
The best-fitted parameters obtained by  $\chi^2$-minimization 
are in good agreement between $^{13}$CO and C$^{18}$O. 
Using $^{13}$CO detected with a higher S/N, the system 
inclination angle is estimated as $i = 26.5^{+2.2}_{-1.7}$ degrees, 
where the uncertainty arises from the error in the stellar mass. 
Note that $i$ is not large enough to yield reasonable constraints on 
both the stellar mass and the inclination \citep{Simon00}. The systemic 
velocity is estimeted to be $3.72 \pm 0.02$~km~s$^{-1}$ in $v_{\mathrm{LSR}}$, 
and the obtained location of the center of mass matches that of 
the compact component of the continuum emission, which most 
likely represents the inner disk. Figure~\ref{fig:pv_major} also 
shows the curves for the Keplerian rotation with the 
parameters adopted in the modeling, i.e., $i=27^{\circ}$, 
$M=2.2~M_{\odot}$ and the systemic velocity of $3.7$ km s$^{-1}$ 
in $v_{\mathrm{LSR}}$. 
No significant deviation from the Keplerian motion was detected within 
the effective resolution of 0.2~km~s$^{-1}$ in our observations.

\clearpage

\section{Details of parameter search for dust distribution}
\label{sec:app_dustdistribution}

In this section, we present the results of parameter search for dust
distribution models.  We vary the parameters for dust distribution 
$\Sigma_{d,0}$, $r_d$ and $w_d$ and look for a set of parameters that
best reproduces the dust continuum observations.  
Here, we exclusively show the results of the north region where dust
emission is optically thick.  For the south region, the disk is
optically thin and the parameter search is straightforward.

For each set of the parameters $(\Sigma_{d,0}, r_d, w_d)$, we calculate
the dust continuum emission and the model images.  
The model images are convolved with the Gaussian beam with the size of
the observations. 
The convolved radial profiles of the surface brightness 
is fitted by the Gaussian function given  
in equation \eqref{eq:Gauss_obs}.  
The derived parameters $(I_p, r_0, w)$ are  
compared with those derived from observations 
$(I_{p,obs}, r_{0,obs}, w_{obs})$.  
Tables \ref{table:dust_varsig}, \ref{table:dust_varrd}, and
\ref{table:dust_varwd} show the results when $\Sigma_{d,0}$, $r_d$, and
$w_d$ are varied, respectively.

The peak brightness of dust emission depends weakly on the assumed peak
surface density.  It is 1.17 and 1.25~Jy/asec$^2$ for 
$\Sigma_{d,0}=0.5$ and $0.7$~g/cm$^2$, respectively, when other
parameters are fixed at $r_d=173$~AU and $w_d=27$~AU (the best-fit
parameters; table \ref{table:dust_varsig}).  
In other words, a factor of 1.4 increase in surface
density results in only $\sim 7$~\% in the surface brightness.  
The peak brightness depends weakly on $r_d$ and $w_d$ as well.  
It is 1.24 and 1.19~Jy/asec$^2$ for $r_d=163$ and 183~AU, 
respectively (table \ref{table:dust_varrd}), meaning that
12~\% change in $r_d$ results in 4~\% change in the peak brightness.
In this case, $r_0$ also changes by $\sim 20$~AU so $r_0$ is
more affected by the change in $r_d$.  When $w_d$
is varied from $22$ to 34~AU, the peak brightness changes 
from 1.12 to 1.28~Jy/asec$^2$ (table \ref{table:dust_varwd}), meaning
that $\sim 50$~\% change in the width of dust distribution results in 
$\sim 14$~\% change in the peak brightness.  In this case, the
width of surface brightness changes by $\sim 20$~\% so it is more
affected than the peak brightness.

\clearpage

\begin{table*}
 \begin{center}
 \begin{tabular}{cc}
   Model Parameters & Model Results \\
  ($\Sigma_{d,0}$~[g/cm$^2$], $r_d$~[AU], $w_d$~[AU]) &
  ($I_{p}$~[Jy/asec$^2$], $r_{0}$~[AU], $w$~[AU]) \\
  \hline
  (0.5, 173, 27) & (1.17, 152, 50)  \\
  (0.6, 173, 27) & (1.20, 153, 50)  \\
  (0.7, 173, 27) & (1.25, 150, 50)  \\
 \end{tabular}
 \end{center}
 \caption{Results of model dust continuum emission of the north profile 
 with different $\Sigma_{d,0}$.  
 The observed parameters are 
 $(I_{p,obs}, r_{0,obs}~[AU], w_{obs}~[AU]) 
 = (1.2~\mathrm{Jy/asec^2}, 152~\mathrm{AU}, 51~\mathrm{AU})$.
 }
 \label{table:dust_varsig}
\end{table*}

\begin{table*}
 \begin{center}
  \begin{tabular}{cc}
   Model Parameters & Model Results  \\
   ($\Sigma_{d,0}$~[g/cm$^2$], $r_d$~[AU], $w_d$~[AU]) &
       ($I_{p}$~[Jy/asec$^2$], $r_{0}$~[AU], $w$~[AU]) \\
   \hline
   (0.6, 163, 27) & (1.24, 141, 51)  \\
   (0.6, 173, 27) & (1.20, 153, 50)  \\
   (0.6, 183, 27) & (1.19, 161, 51)  \\
  \end{tabular}
 \end{center}
 \caption{Results of model dust continuum emission of the north profile 
 with different $r_d$.
 The observed parameters are 
 $(I_{p,obs}, r_{0,obs}~[AU], w_{obs}~[AU]) 
 = (1.2~\mathrm{Jy/asec^2}, 152~\mathrm{AU}, 51~\mathrm{AU})$.
 }
 \label{table:dust_varrd}
\end{table*}

\begin{table*}
 \begin{center}
  \begin{tabular}{cc}
   Model Parameters & Model Results \\
   ($\Sigma_{d,0}$~[g/cm$^2$], $r_d$~[AU], $w_d$~[AU]) &
       ($I_{p}$~[Jy/asec$^2$], $r_{0}$~[AU], $w$~[AU]) \\
   \hline
   (0.6, 173, 22) & (1.12, 154, 47) \\
   (0.6, 173, 27) & (1.20, 153, 50) \\
   (0.6, 173, 34) & (1.28, 146, 56) \\
  \end{tabular}
 \end{center}
 \caption{Results of model dust continuum emission of the north profile 
 with different $w_d$.
 The observed parameters are 
 $(I_{p,obs}, r_{0,obs}~[AU], w_{obs}~[AU]) 
 = (1.2~\mathrm{Jy/asec^2}, 152~\mathrm{AU}, 51~\mathrm{AU})$.
 }
 \label{table:dust_varwd}
\end{table*}


\end{document}